\documentclass[review]{elsarticle}

\usepackage{lineno,hyperref}
\usepackage{graphicx}
\usepackage{dcolumn}
\usepackage{bm}
\usepackage{epstopdf}
\usepackage{color}
\usepackage{textcomp}
\usepackage{amsmath}

\usepackage{booktabs}  
\usepackage{siunitx} 
\modulolinenumbers[5]
\usepackage{multirow}  

\journal{Combustion and Flame}

\begin{document}

\begin{frontmatter}

\title{Burnett-level multi-relaxation-time central-moment discrete Boltzmann modeling of reactive flows}

\author[address1]{Qingbin Wu}

\author[address2,address3]{Chuandong Lin\corref{mycorrespondingauthor1}}
\cortext[mycorrespondingauthor1]{Corresponding author}
\ead{linchd3@mail.sysu.edu.cn}

\author[address1]{Huilin Lai\corref{mycorrespondingauthor2}}
\cortext[mycorrespondingauthor2]{Corresponding author}
\ead{hllai@fjnu.edu.cn}

\address[address1]{School of Mathematics and Statistics, Key Laboratory of Analytical Mathematics and Applications (Ministry of Education), Fujian Key Laboratory of Analytical Mathematics and Applications (FJKLAMA), Center for Applied Mathematics of Fujian Province (FJNU), Fujian Normal University, 350117 Fuzhou, China. }

\address[address2]{Sino-French Institute of Nuclear Engineering and Technology, Sun Yat-sen University, Zhuhai 519082, China.}

\address[address3]{Key Laboratory for Thermal Science and Power Engineering of Ministry of Education, Department of Energy and Power Engineering, Tsinghua University, Beijing 100084, China.}

\begin{abstract}
A multi-relaxation-time central-moment discrete Boltzmann method (CDBM) is developed for compressible reactive flows, incorporating the effects of chemical reactions. The Chapman--Enskog multiscale analysis demonstrates that the model recovers the Burnett equations in the hydrodynamic limit, with tunable specific heat ratios and Prandtl numbers. Within the CDBM framework, a unified Boltzmann equation governs the evolution of hydrodynamic variables, thermodynamic quantities, and higher-order central moments. The collision and reaction term are consistently computed via matrix inversion method. A two-dimensional twenty-five discrete velocities, exhibiting favorable spatial symmetry, is constructed and employed. The model is validated through simulations of the thermal Couette flow, homogeneous chemical reaction, steady detonation wave, and collision of two detonation waves. This work presents a versatile numerical simulation tool capable of addressing complex reactive flows characterized by hydrodynamic and thermodynamic nonequilibrium effects, applicable to both scientific research and engineering practice.

\end{abstract}


\end{frontmatter}


\section{Introduction}

Chemical reactive flows, which entail the intricate coupling between chemical reactions and fluid dynamics, play a foundational role across a wide array of scientific and engineering domains, including combustion technology \cite{Law2006}, detonations \cite{fickett2000detonation}, photocatalysis \cite{corcoran2020photon}, and electrochemical energy storage, etc. \cite{heeschen2024low}. Conventional energy systems, such as coal-fired power plants and oil-based internal combustion engines, are characterized by low thermal efficiency, significant energy dissipation, and high pollutant emissions, thereby posing serious environmental challenges. In particular, the combustion and pyrolysis processes involved in these systems exhibit strong multiscale and multiphysics coupling, making them difficult to simulate accurately using traditional computational approaches and thus limiting their optimization potential \cite{peters2001turbulent}.
In recent years, chemical reactive flow has become increasingly critical for enabling breakthroughs in energy conversion, enhanced energy utilization, and advanced energy storage technologies \cite{heeschen2024low}. Simultaneously, the emergence of clean and renewable energy technologies---such as hydrogen combustion, biomass conversion, and fuel cells---has introduced new opportunities for sustainable development \cite{zheng2023combustion}. However, these technologies are governed by complex chemical kinetics, rapid high-temperature reactions, and turbulent flow phenomena, all of which demand more sophisticated and accurate numerical simulation techniques \cite{oran2001numerical}. 

In essence, chemical reactive flows embody the dynamic coupling of chemical reactions and fluid dynamics, often generating vortex and turbulent structures of different morphologies under flow instability. The processes span multiple spatial and temporal scales, and are accompanied by significant hydrodynamic, thermodynamic, and chemical nonequilibrium effects, presenting complex physicochemical properties \cite{williams2018combustion,nagnibeda2009non}. Due to the continuum medium assumption, traditional macroscopic models, such as the Euler and Navier--Stokes (NS) equations, often fall to accurately describe these multiscale nonequilibrium behaviors; while the applicable scales of microscopic molecular dynamics are limited due to the computational cost \cite{mao2023classical}. Therefore, it is particularly important to find appropriate mesoscopic modeling methods between macroscopic and microscopic.

As a mesoscopic computational method originated from statistical physics, the lattice Boltzmann method (LBM) has made remarkable progress in the application to complex flows \cite{Chai2014PRE,Wei2018AMC}, such as turbulence \cite{Wei2017CF,Zhao2023CF}, heat transfer \cite{Li2016PECS,Li2020CSB,Chiappini2018CoupledLB}, non-Newtonian fluids \cite{Chai2011JNFM}, multiphase flows
\cite{Fei2023JFM,Liang2014PRE,Liang2024POF}, and reactive flows \cite{Succi1997JSC,Yamamoto2002JSP,Chiavazzo2011efficient,Di2012EPL,Wissocq2023JCP,Hosseini2024lattice}. In 1997, Succi et al. constructed the first two-dimensional (2D) LBM to deal with the dynamics of combustion \cite{Succi1997JSC}. In 2002, Yamamoto et al. used a 2D LBM to simulate the combustion process of propane and solved the flow, temperature, and concentration fields with the LB equations \cite{Yamamoto2002JSP}. In 2011, Chiavazzo et al. proposed a simplified LBM for solving the combustion field, and verified the method by a 2D simulation of premixed laminar flames in a hydrogen-air mixture, which achieved an efficient simulation of complex reaction mechanisms \cite{Chiavazzo2011efficient}. In 2012, Di Rienzo et al. extended the LBM to low Mach number reactive flows and successfully simulated the reaction process of hydrogen/air mixtures in a microscale channel, which further broadened the scope of application of the LBM \cite{Di2012EPL}. In 2023, Wissocq et al. developed a hybrid LBM method combining LBM with finite volume algorithms to accurately reconstruct cellular structures in 2D detonations for the first time, breaking the bottleneck of LBM in the simulation of complex structures in detonation waves \cite{Wissocq2023JCP}. In 2024, Hosseini et al. applied the LBM to the field of combustion simulation and proposed a strategy to extend the method to compressible flows \cite{Hosseini2024lattice}. 

Despite great advances in LBM, the focus has usually been on incompressible fluids, with relatively few studies of compressible flows and limited ability to characterize nonequilibrium behaviors. To make up for these deficiencies, researchers have further developed the discrete Boltzmann method (DBM) on the basis of the LBM, which not only inherits the advantages of the LBM in numerical efficiency and structural simplicity, but also is able to describe and quantify the thermodynamic nonequilibrium (TNE) and hydrodynamic nonequilibrium (HNE) effects of the system from the mesoscopic scale, and has gradually been widely used in fluid dynamics research \cite{Gan2013EL,Lai2016PRE, Lin2020CESW,Gan2022JFM,Su2022CTP,Xu2023brief}. 
In 2013, Yan et al. proposed a kinetic to simulate combustion phenomena, analyzing TNE behaviors near von Neumann points \cite{Yan2013PF}. 
In 2014, Lin et al. developed a polar-coordinate DBM for compressible flows to simulate complex phenomena such as chemical reaction-fluid dynamics coupling, steady detonations, implosions, and rarefactions \cite{Lin2014PRE,Lin2014CTP}.
In 2015, to investigate combustion with adjustable specific heat ratios and Prandtl numbers, Xu et al. introduced a 2D multi-relaxation-time (MRT) DBM and incorporated chemical reactions by adding a reaction term to the discrete Boltzmann equation \cite{Xu2015PRE}.
In 2016, Lin et al. proposed a DBM for combustion processes and comparatively analyzed the physical accuracy of the coarse-grained model in nonequilibrium behavior by using different discrete velocity models \cite{Lin2016CNF}. In the same year, Zhang et al. simulated and investigated four types of detonation phenomena with different reaction rates based on the DBM and studied the four types of detonation in terms of hydrodynamic quantities, nonequilibrium quantities and entropy productions \cite{Zhang2016CNF}.
In 2018, Gan et al. developed a cross-scale DBM framework for high-speed compressible flows based on the Bhatnagar--Gross--Krook (BGK) model and constructed a 2D Burnett-level DBM with 26 discrete velocities \cite{Gan2018PRE}. 
In 2019, Lin et al. designed an efficient MRT-DBM capable of handling both stable and unstable supersonic reactive flows, with adjustable specific heat ratios and Prandtl numbers \cite{Lin2019PRE}.
Later, Ji et al. firstly extended the DBM  detonation in three dimensional case \cite{Ji2021AIPA,Ji2022JCP}.
In 2023, Su et al. applied DBM to simulate and analyze non-stationary detonation waves \cite{Su2023CTP}. 
In 2024, Lin et al. proposed a MRT-DBM with a split collision term for reactive flows \cite{Lin2024CTP}. 

Roughly speaking, the DBM can be classified into two categories, i.e., the single-relaxation-time DBM and the MRT-DBM. The BGK model is the most popular single-relaxation-time DBM, which uses a single relaxation parameter to control all thermodynamic processes. Although the BGK-DBM features the simplicity, it lacks flexibility in regulating different kinetic modes and has a limited range of applications, such as a fixed Prandtl number ($\Pr = 1$). 
To overcome this limitation, an efficient method is to use the MRT-DBM \cite{Xu2015PRE,Lin2019PRE,Chen2018POF}.
Although the traditional MRT-DBM allows for an adjustable Prandtl numbers, certain quantities are inevitably lost during the recovery of macroscopic fluid equations via the Chapman--Enskog (CE) expansion, necessitating the inclusion of additional correction terms \cite{Xu2015PRE,Lin2018CNF}.

To solve above issues, a MRT central-moment discrete Boltzmann method (CDBM) has been developed \cite{Su2025central}. The CE 
multiscale analysis demonstrates that CDBM can recover conventional macroscopic fluid equations without the need for additional correction terms.
Moreover, CDBM enables direct quantification of nonequilibrium effects in reactive flows arising from thermal fluctuations.
Accordingly, this paper develops a CDBM targeting the Burnett hierarchy for supersonic compressible reactive flows, with tunable specific heat ratios and Prandtl numbers, and constructs a set of two-dimensional twenty-five discrete velocities (D2V25).
The proposed model recovers the Burnett-level macroscopic equations in the central-moment space via the CE multiscale analysis.
In comparison to the NS-lever CDBM \cite{Su2025central}, the Burnett hierarchy employed incorporates higher-order nonequilibrium effects, enabling the description of stronger nonequilibrium behaviors \cite{Gan2018PRE}. This study provides a powerful kinetic method with a solid theoretical foundation for elucidating the mechanisms of high-order hydrodynamic and thermodynamic nonequilibrium effects in supersonic compressible reactive flows.

The rest of the paper is organized as follows. A detailed description of the CDBM at the Burnett level is presented in Sec. \ref{SecII}. In Sec. \ref{SecIII}, the model is validated against typical benchmarks, i.e., the thermal Couette flow, the homogeneous chemical reaction system, the detonation wave, and the collision of two detonation waves. Finally, conclusions are presented in Sec. \ref{SecIV}.

\section{Central-moment-based discrete Boltzmann model}\label{SecII}

Transitioning from the Boltzmann equation to the DBM involves three key steps \cite{Gan2018PRE,Lin2024CTP,xu2024advances}.

(i) Simplification of the collision operator. The collision term in the Boltzmann equation appears as a complex integral, making direct solution challenging. Thus, simplifying the collision term is essential for practical application of the Boltzmann equation. In this work, we utilize the central-moment MRT collision operator \cite{Su2025central}.

(ii) Discretization of particle velocity space. It is essential to ensure that key physical quantities retain consistency between the continuous and discrete models. Accordingly, moment relations required during discretization are determined using the CE multiscale analysis. To enhance both physical accuracy and numerical efficiency, the matrix inversion method is employed for velocity space discretization. 

(iii) Description of nonequilibrium effects. Nonequilibrium information is extracted through the high-order moments of the deviation between the discrete distribution function and its corresponding equilibrium function. Compared to the first two steps for coarse-grained physical modeling, the third step represents the core objective of DBM. 

\subsection{Reactive discrete Boltzmann equation}

The discrete Boltzmann equation involving chemical reactions is as follows
\begin{equation}
\frac{\partial \mathbf{f}}{\partial t} + \mathbf{v} \cdot  \nabla \mathbf{f} = \mathbf{\Omega } + \mathbf{R}
\tt{,}
\label{BoltzmannEq}
\end{equation}
where $t$ represents the time, $\mathbf{v} = \operatorname{diag} \left(
\begin{array}{llll}
\mathbf{v}_1 & \mathbf{v}_2 & \cdots & \mathbf{v}_N
\end{array} \right)$ the discrete velocity,
$\mathbf{f} = \left(
\begin{array}{llll}
f_1 & f_2 & \cdots & f_N
\end{array} \right)^{\mathrm{T}}$ the discrete distribution function, $\nabla$ the Hamiltonian operator, with the total number $N = 25$ in this work. In addition, $\mathbf{\Omega}$ and $\mathbf{R}$ represent collision term and reaction term, respectively.

The system's density $\rho$, flow velocity $\mathbf{u}$, and energy $E$ are related to the discrete velocity distribution function $f_i$ by the following equations: 
\begin{equation}
\rho = \sum_{i}{f}_{i}
\tt{,}
\end{equation}
\begin{equation}
\rho{\mathbf{u}} = \sum_{i}{f}_{i}{\mathbf{v}}_{i}
\tt{,}
\end{equation}
\begin{equation}
{E} = \frac{\rho}{2}\left[ \left(D+I \right)T+u^2\right] = \frac{1}{2}\sum_{i}{f}_{i}\left( {{\left| {\mathbf{v}}_{i}\right| }^{2} + {\eta }_{i}^{2}}\right)
\tt{.}
\end{equation}
From the above three equations, the temperature can be obtained by
\begin{equation}
T = \frac{{2E - \rho {u^2}}}{{\rho \left( {D + I} \right)}}
\tt{.}
\end{equation}
Clearly, the physical quantities update with the evolution of the discrete velocity distribution function in Eq. (\ref{BoltzmannEq}). 

\subsubsection{Collision term}
The collision term describes the change rate of the distribution function due to molecular collisions and is expressed by
\begin{equation}
\mathbf{\Omega } =  - {\mathbf{C}}^{-1}\mathbf{S}\left( {{\bar{\mathbf{f}}} - {\bar{\mathbf{f}}^{\mathrm{eq}}}}\right)
\tt{,}
\label{CollisionTermSum}
\end{equation}
where ${\bar{\mathbf{f}} }= {\left( \begin{array}{llll} \bar{f_1} & \bar{f_2} & \cdots & \bar f_{25} \end{array}\right) }^{\mathrm{T}}$ and ${\mathbf{\bar f}}^{\mathrm{eq}} =$ ${\left( \begin{array}{llll} \bar {f}_1^{\mathrm{eq}} & \bar{f}_2^{\mathrm{eq}} & \cdots & \bar {f}_{25}^{\mathrm{eq}} \end{array}\right) }^{\mathrm{T}}$ are the central moments of the distribution function and the equilibrium distribution function, respectively.  $\mathbf{S} = \operatorname{diag}\left( \begin{array}{llll} {S}_{1} & {S}_{2} & \cdots & {S}_{25} \end{array}\right)$ is the relaxation factor matrix, which controls the relaxation rate of $\bar{\mathbf{f}}$ toward $\bar {\mathbf{f}}^{\mathrm{eq}}$.  $\mathbf{C}=\left( \begin{array}{llll} \mathbf{C}_{1} & \mathbf{C}_{2} & \cdots & \mathbf{C}_{25} \end{array}\right)^\mathrm{T}$ denotes the transformation matrix between velocity space and central moment space, containing the blocks $\mathbf{C}_i=\left( \begin{array}{llll} {C}_{i1} & {C}_{i2} & \cdots & {C}_{i25} \end{array}\right)$, with elements
${C}_{1i} = 1$,
${C}_{2i} = {v}_{ix}^*$,
${C}_{3i} = {v}_{iy}^*$,
${C}_{4i} = {v}_{i}^{*2} + {\eta }_{i}^{2}$,
${C}_{5i} = {v}_{ix}^{*2}$,
${C}_{6i} = {v}_{ix}^*{v}_{iy}^*$,
${C}_{7i} = {v}_{iy}^{*2}$,
${C}_{8i} = \left( {v}_{i}^{*2} + {\eta }_{i}^{2} \right) {v}_{ix}^*$,
${C}_{9i} = \left( {v}_{i}^{*2} + {\eta }_{i}^{2} \right) {v}_{iy}^*$,
${C}_{10i} = {v}_{ix}^{*3}$,
${C}_{11i} = {v}_{ix}^{*2} {v}_{iy}^*$,
${C}_{12i} = {v}_{ix}^* {v}_{iy}^{*2}$,
${C}_{13i} = {v}_{iy}^{*3}$,
${C}_{14i} = \left( {v}_{i}^{*2} + {\eta }_{i}^{2} \right) {v}_{ix}^{*2}$,
${C}_{15i} = \left( {v}_{i}^{*2} + {\eta }_{i}^{2} \right) {v}_{ix}^* {v}_{iy}^*$,
${C}_{16i} = \left( {v}_{i}^{*2} + {\eta }_{i}^{2} \right) {v}_{iy}^{*2}$,
${C}_{17i} = {v}_{ix}^{*4}$,
${C}_{18i} = {v}_{ix}^{*3} {v}_{iy}^*$,
${C}_{19i} = {v}_{ix}^{*2} {v}_{iy}^{*2}$,
${C}_{20i} = {v}_{ix}^* {v}_{iy}^{*3}$,
${C}_{21i} = {v}_{iy}^{*4}$,
${C}_{22i} = \left( {v}_{i}^{*2} + {\eta }_{i}^{2} \right) {v}_{ix}^{*3}$,
${C}_{23i} = \left( {v}_{i}^{*2} + {\eta }_{i}^{2} \right) {v}_{ix}^{*2} {v}_{iy}^*$,
${C}_{24i} = \left( {v}_{i}^{*2} + {\eta }_{i}^{2} \right) {v}_{ix}^* {v}_{iy}^{*2}$, and
${C}_{25i} = \left( {v}_{i}^{*2} + {\eta }_{i}^{2} \right) {v}_{iy}^{*3}$. The superscript $\rm{T}$ indicates the transpose of a matrix. $\mathbf{C}^{-1}$ is the inverse matrix of $\mathbf{C}$. Furthermore, The inverse matrix $\mathbf{C}^{-1}$ can be computed by using MATLAB.

In order to recover the compressible Burnett equations, the discrete equilibrium distribution function
${\mathbf{f}}^{\mathrm{eq}} = {\left( \begin{array}{llll} {f}_{1}^{\mathrm{eq}} & {f}_{2}^{\mathrm{eq}} & \cdots & {f}_{25}^{\mathrm{eq}} \end{array}\right) }^{\mathrm{T}}$
should satisfy the following relations:
\begin{equation}
\iint {f^{\mathrm{eq}}\Psi d}\mathbf{v}{d\eta } = \mathop{\sum }\limits_{i}{f}_{i}^{\mathrm{eq}}{\Psi }_{i}
\tt{,}
\label{feq_M}
\end{equation}
where $\Psi=1$, $\mathbf{v}^*$, ${\mathbf{v}^* \cdot \mathbf{v}^* + {\eta}^{2}}$, $\mathbf{v}^* \mathbf{v}^*$, $\left({\mathbf{v}^* \cdot \mathbf{v}^* + {\eta }^{2}}\right) \mathbf{v}^*$, $\mathbf{v}^* \mathbf{v}^* \mathbf{v}^* $, $\left( \mathbf{v}^* \cdot \mathbf{v}^* + \eta^2 \right)\mathbf{v}^* \mathbf{v}^*$, $\mathbf{v}^* \mathbf{v}^* \mathbf{v}^* \mathbf{v}^*$, $\left(\mathbf{v}^* \cdot \mathbf{v}^*+\eta ^2 \right)\mathbf{v}^* \mathbf{v}^* \mathbf{v}^* $, accordingly, $\Psi_i=1$, $\mathbf{v}_i^*$, ${\mathbf{v}_i^* \cdot  \mathbf{v}_i^* + {\eta }^{2}}$, $\mathbf{v}_i^* \mathbf{v}_i^*$, $ \left( {\mathbf{v}_i^* \cdot  \mathbf{v}_i^* + {\eta }^{2}}\right) \mathbf{v}_i^*$, $\mathbf{v}_i^* \mathbf{v}_i^* \mathbf{v}_i^*$, $ \left( \mathbf{v}_i^* \cdot  \mathbf{v}_i^*+\eta^2 \right)\mathbf{v}_i^* \mathbf{v}_i^*$, $ \mathbf{v}_i^* \mathbf{v}_i^* \mathbf{v}_i^* \mathbf{v}_i^*$, $ \left( \mathbf{v}_i^* \cdot  \mathbf{v}_i^*+\eta^2 \right)\mathbf{v}_i^* \mathbf{v}_i^* \mathbf{v}_i^*$.
Here $\mathbf{v}^* = \mathbf{v} - \mathbf{u}$ and $\mathbf{v}_i^* = \mathbf{v}_i - \mathbf{u}$ represent the peculiar velocities relative to the flow velocity $\mathbf{u}$, while $\eta$ and $\eta_i$ correspond to vibrational and/or rotational energy. On the left-hand side of Eq.(\ref{feq_M}), $f^{\mathrm{eq}}$ denotes the equilibrium distribution function and takes the form \cite{Lin2017PRE}	
\begin{equation}
{f^{\mathrm{eq}}} = n{\left( {\frac{m}{{2\pi T}}} \right)^{D/2}}{\left( {\frac{m}{{2\pi IT}}} \right)^{1/2}}\exp \left[ { - \frac{{m{{\left| {{\mathbf{v}} - {\mathbf{u}}} \right|}^2}}}{{2T}} - \frac{{m{\eta ^2}}}{{2IT}}} \right]
\tt{,}
\label{feq}
\end{equation}
where $n$ denotes the particle number density, $m=1$ the particle mass, and $\rho = mn$ the particle mass density, $T$ the temperature, $D=2$ the number of spatial dimension, and $I$ the additional degrees of freedom arising from vibration and/or rotation. Substituting Eq. (\ref{feq}) into (\ref{feq_M}) leads to an explicit expression:
\begin{equation}
\bar{\mathbf{f}}^{\mathrm{eq}}=\mathbf{C} \cdot \mathbf{f}^{\mathrm{eq}}
\tt{,}
\end{equation}
where ${\mathbf{\bar f}}^{\mathrm{eq}} =$ ${\left( \begin{array}{llll} \bar {f}_1^{\mathrm{eq}} & \bar{f}_2^{\mathrm{eq}} & \cdots & \bar {f}_{25}^{\mathrm{eq}} \end{array}\right) }^{\mathrm{T}}$ is the column vector, with elements $\bar{f}_{1}^{\mathrm{eq}}=\rho$, $\bar{f}_{2}^{\mathrm{eq}}=0$, $\bar{f}_{ 3}^{\mathrm{eq}}=0$, $\bar{f}_{4}^{\mathrm{eq}}= \left( D+I \right)\rho T$, $\bar{f}_{5}^{\mathrm{eq}}=\rho T$, $\bar{f}_{6}^{\mathrm{eq}}=0$, $\bar{f}_{7}^{\mathrm{eq}}=\rho T$, $\bar{f }_{8}^{\mathrm{eq}}=0$, $\bar{f}_{9}^{\mathrm{eq}}=0$, $\bar{f}_{10}^{\mathrm{eq}}=0$, $\bar{f}_{11}^{\mathrm{eq}}=0$, $\bar{f}_{12}^{\mathrm{eq}}=0$, $\bar{f}_{13}^{\mathrm{eq}}=0$, $\bar{f }_{14}^{\mathrm{eq}}= \left( D+I+2 \right)\rho T^2$, $\bar{f}_{15}^{\mathrm{eq}}=0$, $\bar{f}_{16}^{\mathrm{eq}}= \left( D+I+2\right)\rho T^2$, $\bar{f}_{17}^{\mathrm{eq}}=3 \rho T^2$, $\bar{f}_{18}^{\mathrm{eq}}=0$, $\bar{f}_{19}^{\mathrm{eq}}=\rho T^2$, $\bar{f}_{20}^{\mathrm{eq}}=0$, $\bar{f}_{21}^{\mathrm{eq}}=3\rho T^2$, $\bar{f}_{22}^{\mathrm{eq}}=0$,  $\bar{f}_{23}^{\mathrm{eq}}=0$, $\bar{f}_{24}^{\mathrm{eq}}=0$, and $\bar{f}_{25}^{\mathrm{eq}}=0$. Thus, as long as the matrix $\mathbf{C}$ is invertible, the discrete equilibrium distribution function $f_i^{\mathrm{eq}}$ can be obtained via
\begin{equation}
\mathbf{f}^{\mathrm{eq}}=\mathbf{C}^{-1} \cdot \bar{\mathbf{f}}^{\mathrm{eq}}
\tt{.}
\end{equation}

\subsubsection{Reaction term}
The reaction term is the change rate of the distribution function due to the chemical reaction. In this work, the chemical reaction process is described using a two-step chemical reaction model \cite{Ng2005numerical}: the first step represents the ignition process within the thermally neutral induction zone and is described by the variable $\xi$, while the second step represents the process of rapid energy release within the chemically exothermic zone and is described by the variable $\lambda$. Specifically, the variables $\xi$ and $\lambda$ have time change rates as bellow:
\begin{equation}
\xi^{\prime} = Hk_I\exp \left[ E_I\left( T_s^{-1}-T^{-1 }\right)\right]
\tt{,}
\end{equation}
\begin{equation}
\lambda^{\prime} = \left( 1-H \right)k_R\left( 1-\lambda \right)\exp \left( -E_RT^{-1} \right)
\tt{,}
\end{equation}
where $H$ is the step function of $\xi$. For $\xi < 1$, $H=1$; for $\xi \geq 1$, $H=0$. The parameters $k_I$ and $k_R$ are the rate constants for the ignition and reaction processes, $E_I$ and $E_R$ are the corresponding activation energies, and $T_s$ is the post-wave temperature.

Under the condition that the chemical reaction proceeds more slowly than the TNE relaxation process, the reaction term can be expressed mathematically as follows \cite{Xu2015PRE}:
\begin{equation}
R = \frac{-\left( {1 + D}\right) {IT} + I{\left| \mathbf{v} - \mathbf{u}\right| }^{2} + {\eta }^{2}}{{2I}{T}^{2}}{f}^{\mathrm{{eq}}}{T}^{\prime }
\tt{,}
\label{Reaction term}
\end{equation}
where ${T}^{\prime } = {2Q}{\lambda }^{\prime }/\left( {D + I}\right)$ denotes the change rate of temperature, $Q$ the chemical heat release per unit mass of fuel, and $\lambda$ the mass fraction of chemical product.

Accordingly, the reaction term $R_i$ in discrete form needs to satisfy the following relations
\begin{equation}
\iint {R\Psi d}\mathbf{v}{d\eta } = \mathop{\sum }\limits_{i}{R}_{i}{\Psi }_{i}
\tt{,}
\label{R_M}
\end{equation}
where $\Psi$ and ${\Psi }_{i}$ have the same elements as in Eq. (\ref{feq_M}). Substituting Eq. (\ref{Reaction term}) into Eq. (\ref{R_M}) yields
\begin{equation}
\bar{\mathbf{R}} = \mathbf{C} \cdot \mathbf{R}
\tt{,}
\end{equation}
then the reaction term can be obtained by
\begin{equation}
\mathbf{R} = \mathbf{C}^{-1} \cdot \bar{\mathbf{R}}
\tt{,}
\end{equation}
where $\mathbf{R} = {\left( \begin{array}{llll} {R}_{1} & {R}_{2} & \cdots & {R}_{25} \end{array}\right) }^{\mathrm{T}}$ and $\bar {\mathbf{R}} = {\left( \begin{array}{llll} \bar{{R}}_{1} & \bar{{R}}_{2} & \cdots & \bar{{R}}_{25} \end{array}\right) }^{\mathrm{T}}$ are the expressions for the reaction term in velocity space and central moment space, respectively. The elements are specified as $\bar{R}_{1}=0$, $\bar{R}_{2}=0$, $\bar{R}_{3}=0$, $\bar{R}_{4}=2\rho \lambda^{\prime} Q$, $\bar{R}_{5}=2\rho \lambda^{\prime} Q/\left( D+I \right)$, $\bar{R}_{6}=0$, $\bar{R}_{7}=2\rho \lambda^{\prime} Q/\left( D+I \right)$, $\bar{R}_{8}=0$, $\bar{R}_ {9}=0$, $\bar{R}_{10}=0$, $\bar{R}_{11}=0$, $\bar{R}_{12}=0$, $\bar{R}_{13}=0$, $\bar{R}_{14}=2\rho \lambda^{\prime} Q T\left( D+I+2 \right) /\left( D+I \right)$, $\bar{R}_{15}=0$, $\bar{R}_{16}=2\rho \lambda^{\prime} Q T\left( D+I+2 \right)/\left( D+I \right)$, $\bar{R}_{17}=12\rho \lambda^{\prime} Q T/\left( D+I \right)$, $\bar{R}_{18}=0$, $\bar{R}_{19}=4\rho \lambda^{\prime} Q T/\left( D+I \right)$, $\bar{R}_{20}=0$, $\bar{R}_{21}=12\rho \lambda^{\prime} Q T/\left( D+I \right)$, $\bar{R}_{22}=0$, $\bar{R}_{23}=0$, $\bar{R}_{24}=0$, and $\bar{R}_{25}=0$.

\subsection{Discretization of velocity}

\begin{figure}
	\begin{center}
		\includegraphics[width=0.45\textwidth]{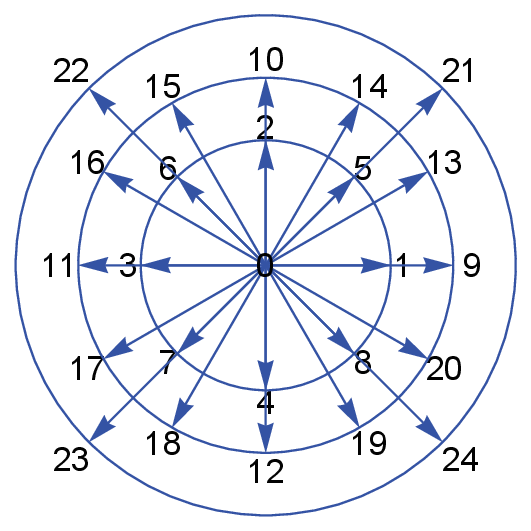}
	\end{center}
	\setlength{\abovecaptionskip}{-0.4cm}
	\caption{Sketch of the discrete velocities.}
	\label{Fig01}
\end{figure}

As shown in Fig. \ref{Fig01}, a model of two-dimensional twenty-five velocities (D2V25) is constructed, with the discrete velocities expressed by
\begin{equation}
{{\mathbf{v}}_i} = \left\{ {\begin{array}{*{20}{l}}
	{0,}&{i = 0,} \\
	{{\text{cyc:}}{v_a}\left( { \pm 1,0} \right),}&{1 \leq i \leq 4,} \\
	{{v_a}\left( { \pm \frac{{\sqrt 2 }}{2} , \pm \frac{{\sqrt 2 }}{2}} \right)\tt{,}}&{5 \leq i \leq 8,} \\
	{{\text{cyc:}}{v_b}\left( { \pm 1,0} \right),}&{9 \leq i \leq 12,} \\
	{{\text{cyc:}}{v_b}\left( { \pm \frac{{\sqrt 3 }}{2} , \pm 0.5} \right),}&{13 \leq i \leq 20,} \\
	{{v_c}\left( { \pm \frac{{\sqrt 2 }}{2} , \pm \frac{{\sqrt 2 }}{2}} \right),}&{21 \leq i \leq 24\tt{,}}
	\end{array}} \right.
\end{equation}
where $v_a$, $v_b$, and $v_c$ are adjustable parameters controlling the size of discrete velocities, and ``cyc" denotes the cyclic substitution, which generates the corresponding vectors by cyclically rearranging the numbers and symbols in the velocity vectors. In order to characterize the internal energy corresponding to molecular rotation and vibration, we also introduce the symbol $\eta_i$ defined as
\begin{equation}
{\eta _i} = \left\{ {\begin{array}{*{20}{l}}
	0,&{i = 0,} \\
	{{\eta _a}},&{1 \leq i \leq 8,} \\
	{{\eta _b}},&{9 \leq i \leq 12,} \\
	{{\eta _c}},&{13 \leq i \leq 24\tt{.}}
	\end{array}} \right.
\end{equation}
Correspondingly, $\eta_a$, $\eta_b$, and $\eta_c$ are also adjustable. It is worth noting that, in order to improve numerical stability and numerical accuracy,  the values of $v_a$, $v_b$, $v_c$ can be given around the flow velocity $\mathbf{u}$ and the sound velocity $v_s=\sqrt{\gamma T/m}$, with $\gamma=(D+I+2)/(D+I)$ the specific heat ratio. Meanwhile, according to the energy equalization theorem, the additional internal energy is approximately $ \frac{1}{2}m{\bar{\eta }}^{2} = \frac{1}{2}IT $, thus one (another) parameter of $\eta_a$, $\eta_b$, and $\eta_c$ should be less than (greater than) $ {\bar{\eta }} = \sqrt{ IT/m} $.

\subsection{Hydrodynamic Modeling and Nonequilibrium Quantities}

The CE multiscale analysis demonstrates that CDBM can recover the set of macroscopic hydrodynamic equations at the Burnett level containing chemical reactions in the continuum limit condition:
\begin{equation}
\frac{{\partial \rho }}{{\partial t}} + \nabla  \cdot  \left( {\rho \mathbf{u}}\right) = 0
\tt{,}
\label{mass}
\end{equation}
\begin{equation}
\frac{{\partial }}{{\partial t}}\left( {\rho \mathbf{u}}\right) + \nabla  \cdot  \left( {\rho \mathbf{{uu}} + p\mathbf{I} + {\mathbf{\Delta }}_{2}^{ * }}\right) = 0
\tt{,}
\label{momentum}
\end{equation}
\begin{equation}
\frac{{\partial E}}{{\partial t}} + \nabla  \cdot  \left\lbrack  {\left( {E + p}\right) \mathbf{u} + {\mathbf{\Delta }}_{2}^{ * } \cdot  \mathbf{u} + {\mathbf{\Delta }}_{3,1}^{ * }}\right\rbrack =  \rho \lambda^{\prime}Q
\tt{,}
\label{energy}
\end{equation}
where $p=\rho T$ denotes the pressure, $E=e+ \rho u^2/2$ is the total energy density, and $e=\left( D+I \right)\rho T/2$ is the internal energy density. 

It is important to note that Eq. (\ref{feq_M}) is prerequisites for deriving the Burnett-level equations from the CDBM.
According to the conservation law, when $\Psi=1$, $\mathbf{v}^*$, ${\mathbf{v}^* \cdot \mathbf{v}^* + {\eta}^{2}}$, the equation still holds when $f_i^{\mathrm{eq}}$ in Eq. (\ref{feq_M}) is replaced by $f_i$, yielding three equations corresponding to the conservation of mass, momentum, and energy, respectively. In contrast, when $\Psi=\mathbf{v}^* \mathbf{v}^*$, $\left({\mathbf{v}^* \cdot \mathbf{v}^* + {\eta }^{2}}\right) \mathbf{v}^*$, $\mathbf{v}^* \mathbf{v}^* \mathbf{v}^* $, $\left( \mathbf{v}^* \cdot \mathbf{v}^* + \eta^2 \right)\mathbf{v}^* \mathbf{v}^*$,$\mathbf{v}^* \mathbf{v}^* \mathbf{v}^* \mathbf{v}^*$, $\left(\mathbf{v}^* \cdot \mathbf{v}^*+\eta ^2 \right)\mathbf{v}^* \mathbf{v}^* \mathbf{v}^* $, if $f_i^{\mathrm{eq}}$ is replaced by $f_i$ in Eq. (\ref{feq_M}), the moments of the same order of $f_i^{\mathrm{eq}}$ and $f_i$ may deviate before and after the substitution. These deviations quantify the degree of departure from TNE from multiple perspectives. Mathematically, we define
\begin{equation}
{\mathbf{\Delta }}_{m,n}^* = \sum_i \left(\frac{1}{2}\right)^{1-\delta_{m,n}} (f_i - f_i^{eq}) \overbrace{\mathbf{v}_i^* \mathbf{v}_i^* \cdots \mathbf{v}_i^*}^{n} (\mathbf{v}_i^{*2} + \eta_i^{2})^{\frac{m-n}{2}}
\tt{,}
\label{delta}
\end{equation}
where $\mathbf{\Delta}_{m,n}^*$ denotes a $m$th-order tensor contracted to a $n$th-order one, and $\delta_{m,n}$ is the Kronecker delta function. When $m=n$, $\mathbf{\Delta}_{m,n}^*$ is abbreviated as $\mathbf{\Delta}_m^*$. Physically, $\mathbf{\Delta}_{m,n}^*$ characterizes TNE due to thermal fluctuations arising from the irregular motion of molecules. Specifically,
\begin{equation}
\mathbf{\Delta}_2^* = \sum_i (f_i - f_i^{\mathrm{eq}}) \mathbf{v}_i^* \mathbf{v}_i^*
\tt{,}
\label{delta_2^*}
\end{equation}
\begin{equation}
\mathbf{\Delta}_{3,1}^* = \sum_i \frac{1}{2} (f_i - f_i^{\mathrm{eq}}) (\mathbf{v}_i^* \cdot \mathbf{v}_i^* + \eta_i^2) \mathbf{v}_i^*
\tt{,}
\label{delta_31^*}
\end{equation}
\begin{equation}
\mathbf{\Delta}_3^* = \sum_i (f_i - f_i^{\mathrm{eq}}) \mathbf{v}_i^* \mathbf{v}_i^* \mathbf{v}_i^*
\tt{,}
\label{delta_3^*}
\end{equation}
\begin{equation}
\mathbf{\Delta}_{4,2}^* = \sum_i \frac{1}{2} (f_i - f_i^{\mathrm{eq}}) (\mathbf{v}_i^* \cdot \mathbf{v}_i^* + \eta_i^2) \mathbf{v}_i^* \mathbf{v}_i^*
\tt{,}
\label{delta_42^*}
\end{equation}
\begin{equation}
\mathbf{\Delta}_4^* = \sum_i (f_i - f_i^{\mathrm{eq}}) \mathbf{v}_i^* \mathbf{v}_i^* \mathbf{v}_i^* \mathbf{v}_i^*
\tt{,}
\label{delta_4^*}
\end{equation}
\begin{equation}
\mathbf{\Delta}_{5,3}^* = \sum_i \frac{1}{2} (f_i - f_i^{\mathrm{eq}}) (\mathbf{v}_i^* \cdot \mathbf{v}_i^* + \eta_i^2) \mathbf{v}_i^* \mathbf{v}_i^* \mathbf{v}_i^*
\tt{.}
\label{delta_53^*}
\end{equation}
$\mathbf{\Delta}_2^*$ corresponds to the viscous stress tensor and the nonorganized energy, $\mathbf{\Delta}_{3,1}^*$ and $\mathbf{\Delta}_3^*$ refer to the flux of nonorganized energy, $\mathbf{\Delta}_{4,2}^*$ represents the flux of nonorganized energy flux, and $\mathbf{\Delta}_4^*$ and $\mathbf{\Delta}_{5,3}^*$ represent higher-order nonequilibrium quantities. 

It is worth noting that previous DBMs often introduce correction term in the collision operator to restore macroscopic relations lost in coarse-grained modeling. In contrast, the CDBM can recover the NS and Burnett level equations via the CE expansion, without requiring additional correction term for the collision operator. This makes CDBM both simpler and more computationally efficient than traditional DBMs. This model employs a matrix inversion approach to compute collision and reaction term, enhancing computational efficiency and numerical accuracy, while also facilitating implementation. Additionally, temporal and spatial derivatives in Eq. (\ref{BoltzmannEq}) are discretized using the second-order Runge--Kutta method and the second-order non-oscillatory, non-free-parameter dissipation finite difference scheme \cite{Zhang1991NND}, respectively.

\subsection{Boundary condition}

Boundary conditions are a critical aspect of numerical simulations of fluid dynamics. To ensure both accuracy and stability, it is crucial to select appropriate boundary conditions based on the specific physical scenario. The DBMs employed in our work support a variety of boundary condition implementations. Among them, the specular reflection boundary condition is a commonly used method,  especially for simulating flows interacting with smooth solid walls. This boundary condition assumes that incident particles reflect off the wall such that the angle of reflection equals the angle of incidence. Specifically, the velocity component normal to the wall is reversed upon reflection, while the tangential component remains unchanged. 

\begin{figure}
	\begin{center}
		\includegraphics[width=0.95\textwidth]{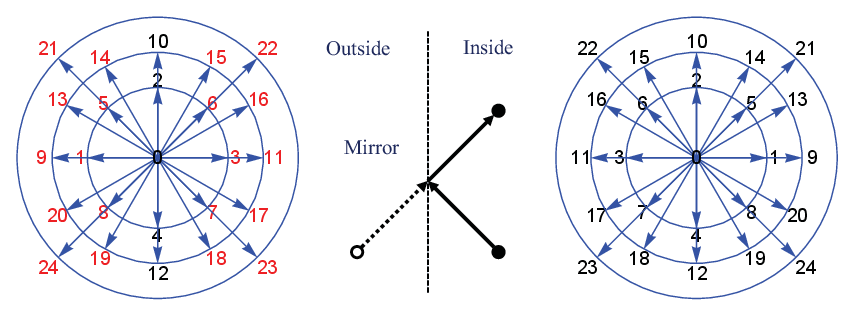}
	\end{center}
	\setlength{\abovecaptionskip}{-0.4cm}
	\caption{Specular reflection boundary in the $x$-direction. }
	\label{Fig02}
\end{figure}

\begin{figure}
	\begin{center}
		\includegraphics[width=0.4\textwidth]{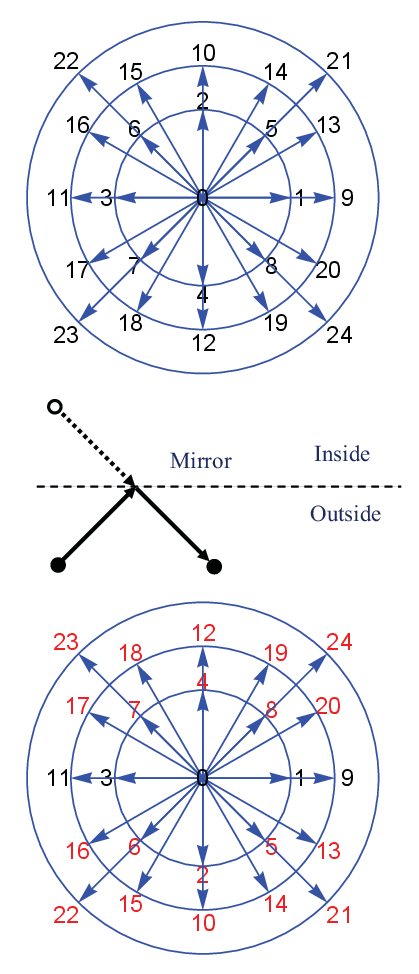}
	\end{center}
	\setlength{\abovecaptionskip}{-0.4cm}
	\caption{Specular reflection boundary in the $y$-direction.}
	\label{Fig03}
\end{figure}

As shown in Fig. \ref{Fig02}, if the wall's normal vector is aligned with the $x$-axis, only the horizontal velocity is reversed, while the other velocities remain constant. Specifically, the reflected discrete distribution functions at the left and right boundaries are expressed as~\cite{Zhang2018CTP}
\begin{equation}
\left\{ \begin{array}{l}
{f_{i,k}}\left( {{v_{ix}}, {v_{iy}}} \right) = {f_{i,1-k}}\left( { - {v_{ix}}, {v_{iy}}} \right), k\le 0 \tt{,}  \hfill \\
{f_{i,k}}\left( {{v_{ix}}, {v_{iy}}} \right) = {f_{i,2{N_x}+1-k}}\left( { - {v_{ix}}, {v_{iy}}} \right), k \ge {N_x}+1 \tt{,} \hfill \\
\end{array}  \right.
\label{boundary_$x$}
\end{equation}
where $f_{i,k}$ denotes the discrete distribution function at the $k$-th ghost grid layer, and $N_x$ represents the total number of grid layers in the $x$-direction. To accommodate the second-order non-oscillatory, non-free-parameter dissipative scheme used for spatial discretization, two layers of ghost cells are introduced at each boundary. Specifically, the ghost layers are located at $k=-1$ and $k=0$ on the left boundary, and at $k=N_x+1$ and $k=N_x+2$ on the right boundary. Additionally, as shown in Fig. \ref{Fig03}, the discrete distribution functions on the ghost points in the $y$-direction can be obtained by using the specular reflection boundary condition in a similar way. In sum, the relations between the distribution functions inside and outside the boundaries are listed in Table \ref{TableI}. 

\begin{table}[h]
	\centering
	\renewcommand{\arraystretch}{0.8}
	\setlength{\tabcolsep}{0.3pt} %
	\begin{tabular}{|c|c|*{25}{>{\centering\arraybackslash}p{0.39cm}|}}
		\hline
		\multicolumn{2}{|c|}{Location} & \multicolumn{25}{c|}{Index of discrete velocities $i$} \\ \hline
		\multicolumn{2}{|c|}{Inside} & 0 & 1 & 2 & 3 & 4 & 5 & 6 & 7 & 8 & 9 & 10 & 11 & 12 & 13 & 14 & 15 & 16 & 17 & 18 & 19 & 20 & 21 & 22 & 23 & 24 \\ \hline
		\multicolumn{2}{|c|}{Outside ($x$)} & 0 & \textcolor{red}{3} & 2 & \textcolor{red}{1} & 4 & \textcolor{red}{6} & \textcolor{red}{5} & \textcolor{red}{8} & \textcolor{red}{7} & \textcolor{red}{11} & 10 & \textcolor{red}{9} & 12 & \textcolor{red}{16} & \textcolor{red}{15} & \textcolor{red}{14} & \textcolor{red}{13} & \textcolor{red}{20} & \textcolor{red}{19} & \textcolor{red}{18} & \textcolor{red}{17} & \textcolor{red}{22} & \textcolor{red}{21} & \textcolor{red}{24} & \textcolor{red}{23} \\ \hline
		\multicolumn{2}{|c|}{Outside ($y$)} & 0 & 1 & \textcolor{red}{4} & 3 & \textcolor{red}{2} & \textcolor{red}{8} & \textcolor{red}{7} & \textcolor{red}{6} & \textcolor{red}{5} & 9 & \textcolor{red}{12} & 11 & \textcolor{red}{10} & \textcolor{red}{20} & \textcolor{red}{19} &  \textcolor{red}{18} & \textcolor{red}{17} & \textcolor{red}{16} & \textcolor{red}{15} & \textcolor{red}{14} & \textcolor{red}{13} & \textcolor{red}{24} & \textcolor{red}{23} & \textcolor{red}{22} & \textcolor{red}{21} \\ \hline
	\end{tabular}
	\caption{Relationship between the discrete distribution functions inside and outside the specular boundaries in the $x$ and $y$ directions.}
	\label{TableI}
\end{table}

\section{Validation and verification}\label{SecIII}

This section presents several benchmark cases to evaluate the effectiveness and accuracy of the proposed CDBM for compressible fluids. (i) Thermal Couette flow is employed to verify the suitability of CDBM for fluid flows with adjustable Prandtl numbers and specific heat ratios. (ii) A homogeneous chemical reaction system is considered to verify the CDBM's ability to naturally couple chemical reactions with physical fields. (iii) A one-dimensional steady-state detonation wave is simulated to verify the CDBM's ability to capture supersonic reactive flows with strong compressible effects. (iv) Finally, a two-dimensional benchmark case involving the collision of two detonation waves is simulated. It should be mentioned that the CDBM with D2V16 at the NS level \cite{Su2025central} is compared with the CDBM with D2V25 at the Burnett level in this section.

\subsection{Thermal Couette flow}

Thermal Couette flow refers to the motion of a viscous fluid confined between two parallel plates undergoing relative shear. It serves as a classical benchmark for evaluating compressible flow models in which viscosity and thermal conductivity are dominant. This subsection presents the thermal Couette flow simulation to demonstrate the applicability of the CDBM to thermal flows with adjustable specific heat ratios and Prandtl numbers. The initial physical field is set as follows: $\rho_0 = 1.0$, $T_0 = 1.0$, and $\mathbf{u}_0 = 0.0$. The lower wall is stationary and keeps at a constant temperature of $T_1 = 1.0$, while the upper plate moves to the right at $u_2 = 0.1$ and is held at $T_2 = 1.0$. Viscous shear stress transfers momentum to the fluid and changes the flow velocity distribution. The distance between the two plates is $H = 0.3$, the time step $\Delta t = 1.0 \times 10^{-4}$, the spatial step $\Delta x = \Delta y = 1.0 \times 10^{-3}$, with the mesh grids ${N_x} \times {N_y} = 1 \times 300$. Periodic boundary conditions are applied in the left and right boundaries, and second-order extrapolated boundary conditions are employed at the top and bottom boundaries.

\begin{figure}
	\begin{center}
		\includegraphics[bbllx=0pt,bblly=0pt,bburx=519pt,bbury=673pt,width=0.47
		\textwidth]{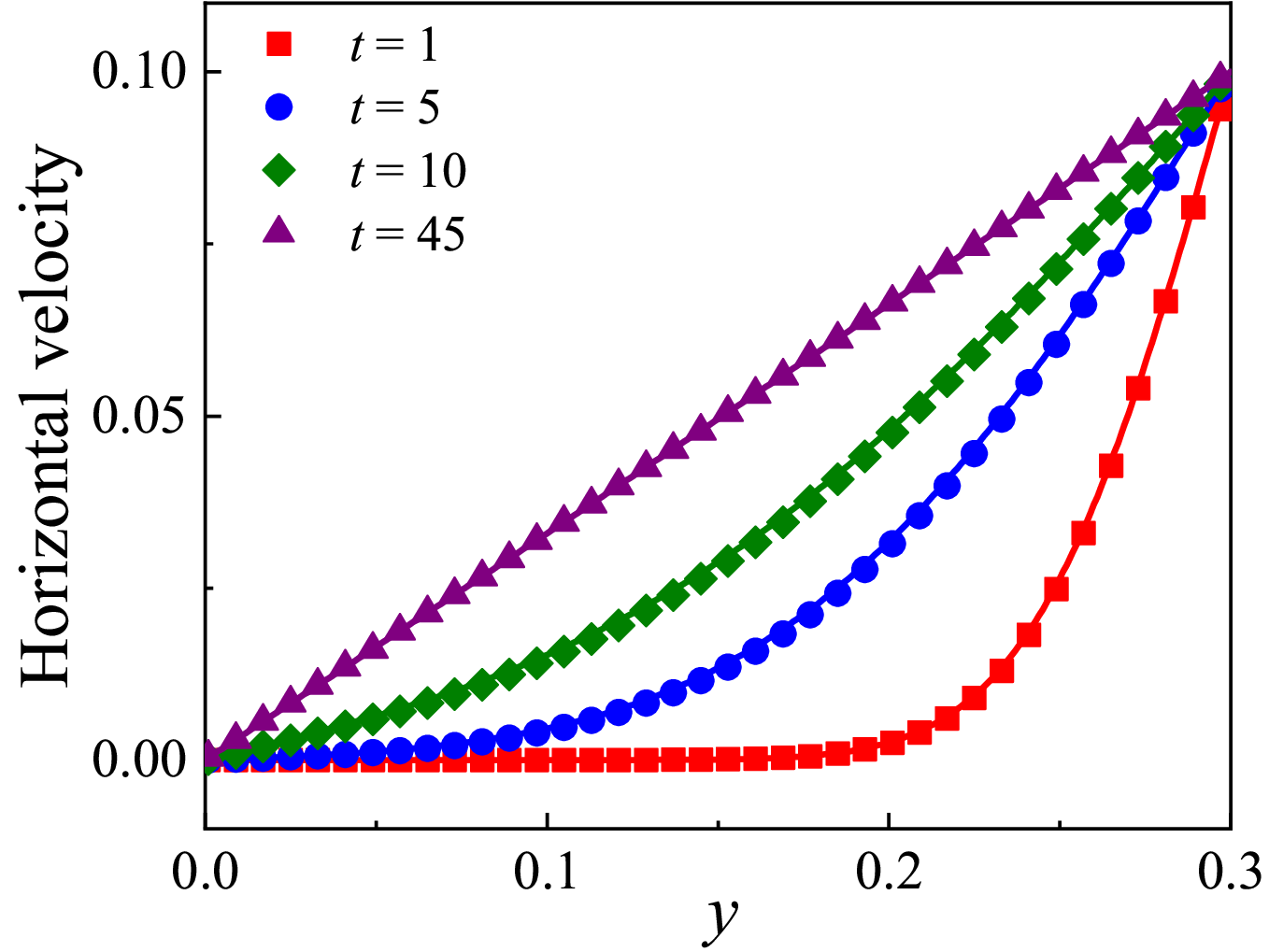}
	\end{center}
	\setlength{\abovecaptionskip}{-0.4cm}
	\caption{Distribution of horizontal velocities along the $y$-direction during the evolution of the thermal Couette flow at different moments with $\gamma=1.5$ and $\Pr=1$. Symbols indicate CDBM results and lines represent exact solutions.}
	\label{Fig04}
\end{figure}

Figure \ref{Fig04} shows the horizontal velocity versus $y$ for the case of $\gamma=1.5$ and $\Pr=1$ when $t$ is taken as $t=1$, $5$, $10$ and $45$, respectively. The symbols represent the simulation results and the lines denote the analytical solutions
\begin{equation}
u = \frac{y}{H}{u}_{0} + \frac{2}{\pi}{u}_{0}\mathop{\sum }\limits_{{n = 1}}^{\infty }\left\lbrack {\frac{{\left( -1\right) }^{n}}{n}\exp \left( {-{n}^{2}{\pi }^{2}\frac{\mu t}{\rho {H}^{2}}}\right) \sin \left( \frac{n\pi y}{H}\right) }\right\rbrack
\tt{.}
\label{u-y}
\end{equation}
Obviously, the simulation results of CDBM coincide with the analytical solutions. 

\begin{figure}
	\begin{center}
		\includegraphics[width=1.0\textwidth]{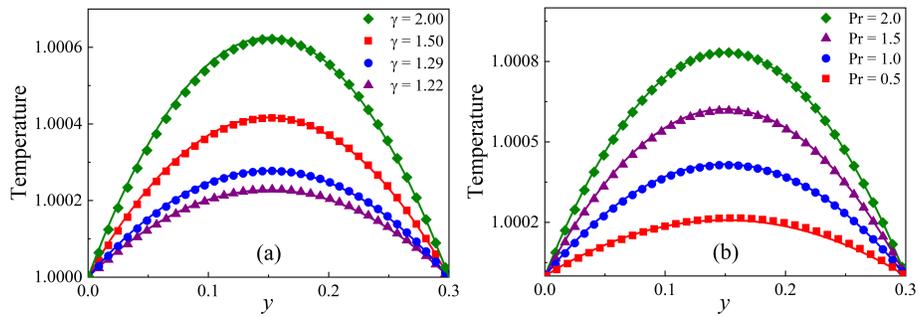}
	\end{center}
	\setlength{\abovecaptionskip}{-0.4cm}
	\caption{(a) Temperature distribution along the $y$-direction for different specific heat ratios with $\Pr=1$. (b) Temperature distribution along the $y$-direction for different Prandtl number cases with $\gamma=1.5$. Symbols indicate CDBM results and lines represent exact solutions.}
	\label{Fig05}
\end{figure}

Theoretically, when the thermal Couette flow reaches equilibrium, the temperature distribution in the $y$-direction follows
\begin{equation}
T = {T}_{1} + \left( {{T}_{2} - {T}_{1}}\right) \frac{y}{H} + \frac{\mu }{2\kappa }{u}_{0}^{2}\frac{y}{H}\left( {1 - \frac{y}{H}}\right),
\label{T-y}
\end{equation}
where $\mu$ and $\kappa$ represent the dynamic viscosity and thermal conductivity, respectively. When $S_5 = S_6 = S_7 = S_\mu$ and $S_8 = S_9 = S_\kappa $, the dynamic viscosity and thermal conductivity are given by $\mu = p/S_{\mu}$ and $\kappa = (D+I+2)p/(2S_{\kappa})$, respectively. 

Figure \ref{Fig05} (a) displays the temperature distributions at different specific heat ratios in the case of $\Pr=1$ when the thermal Couette flow reaches equilibrium, with $\gamma=1.22$, $1.29$, $1.50$, and $2.00$, respectively. The symbols denote the CDBM results, while the lines represent the exact solutions in Eq. (\ref{T-y}). The simulation results are in good agreement with the analytical solutions. Therefore, the CDBM can accurately capture flow fields with different specific heat ratios during the evolution of the thermal Couette flow. Similarly, Fig. \ref{Fig05} (b) illustrates the temperature distribution along the $y$-direction during the evolution of thermal Couette flow for various Prandtl numbers with $\gamma=1.5$. The relaxation parameters are $S_{\kappa} = 1.0 \times 10^3$, with $S_{\mu} = 2.0 \times 10^3$, $1.0 \times 10^3$, $6.7 \times 10^2$, and $5.0 \times 10^2$ respectively corresponding to Prandtl numbers $\Pr = 0.5$, $1.0$, $1.5$, and $2.0$, respectively. Obviously, our simulation results agree well with the analytical solutions for various Prandtl numbers. Consequently, the CDBM is applicable to flow fields with various Prandtl numbers. 

\begin{figure}
	\begin{center}
		\includegraphics[width=0.5\textwidth]{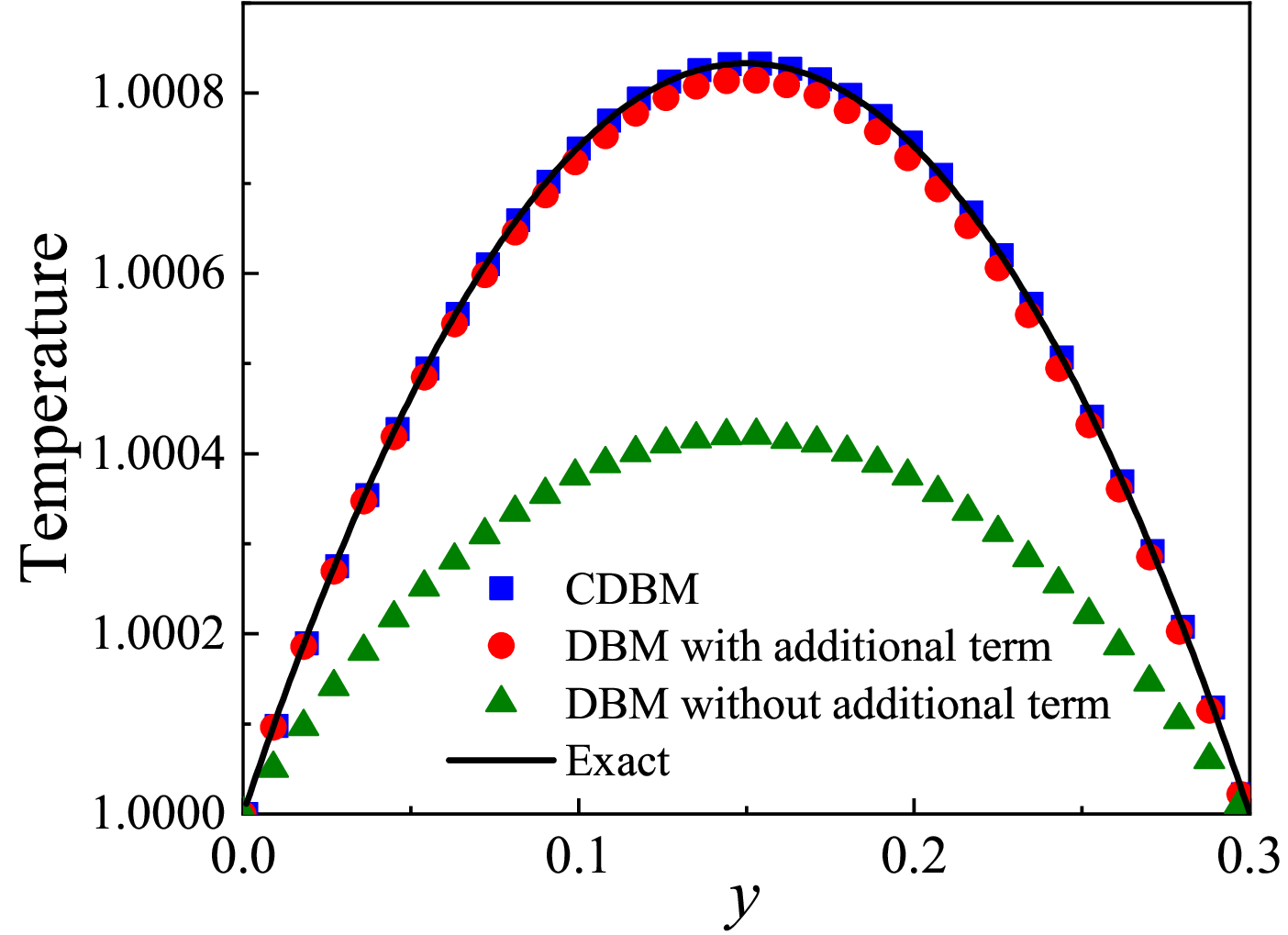}
	\end{center}
	\setlength{\abovecaptionskip}{-0.4cm}
	\caption{Temperature distribution along the $y$-direction for Prandtl number $\Pr = 2$. Squares, circles, and triangles denote the results for the CDBM, the DBM with additional term \cite{Lin2019PRE}, and the DBM without additional term \cite{Chen2011CTP}, respectively. The solid line denotes the exact solutions.}
	\label{Fig06}
\end{figure}

As mentioned above, the CDBM is simpler and more efficient than the traditional MRT-DBM because the CDBM does not require the additional term to correct the collision term. To demonstrate the superior performance of the present model, a comparative analysis with existing models is conducted. Figure \ref{Fig06} shows the temperature distribution along the $y$-direction for $\Pr = 2$, computed using different models: the CDBM (Squares), DBM with additional term (circles) \cite{Lin2019PRE}, and DBM without additional term (triangles) \cite{Chen2011CTP}, respectively. The solid line represents the exact solutions. It is clear that the results of the DBM without the additional term deviate significantly from the exact solutions and the results of the DBM with the additional term are closer to the exact solutions, while the CDBM results show excellent agreement with the exact solutions. Thus, the CDBM exhibits superior performance compared to previous DBMs, particularly when high Prandtl numbers are involved.

\begin{figure}
	\begin{center}
		\includegraphics[width=0.98 \textwidth]{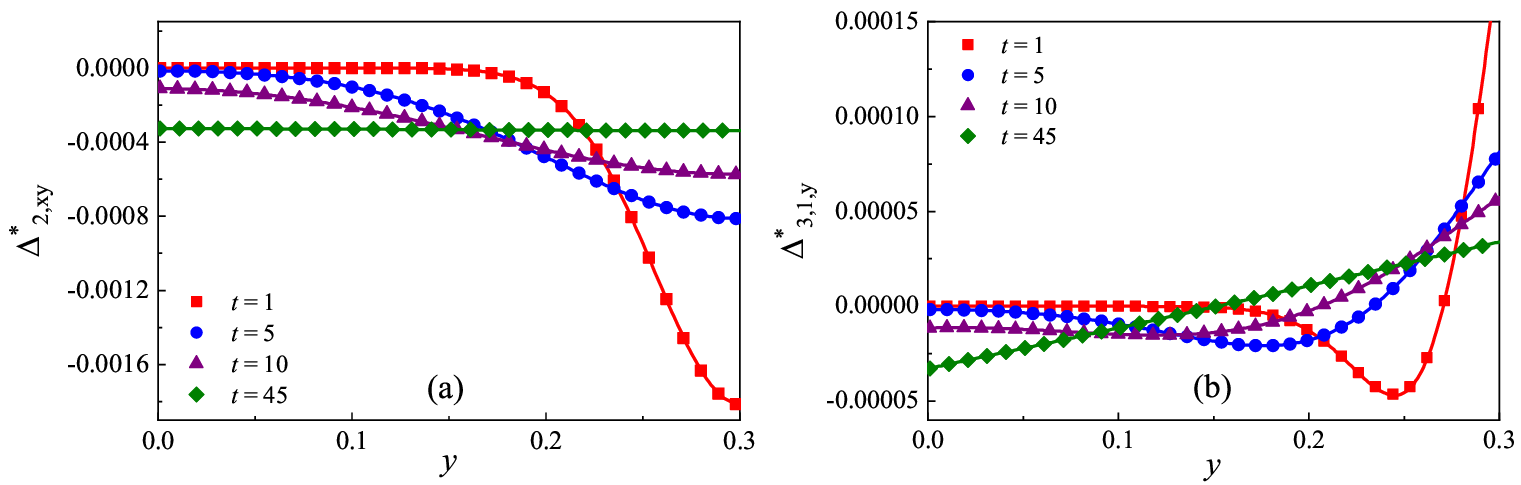}
	\end{center}
	\setlength{\abovecaptionskip}{-0.4cm}
	\caption{Vertical distributions of $\Delta_{2,xy}^*$ and $\Delta_{3,1,y}^*$ at time instants $t=1.0$, $5.0$, $10.0$ and $45.0$, respectively. Symbols indicate CDBM results and solid lines indicate the corresponding theoretical solutions.}
	\label{Fig07}
\end{figure}

To demonstrate that the CDBM can dynamically capture nonequilibrium effects during the iteration process, we examined the nonequilibrium quantities $\Delta_{2,xy}^*$ and $\Delta_{3,1,y}^*$ for $\gamma = 1.5$ and $\Pr=1$. Figure \ref{Fig07} presents the results for the CDBM at $t = 1.0$, $5.0$, $10.0$, and $45.0$, respectively. The symbols represent the results of the CDBM, while the lines correspond to the analytic solutions. By the CE multiscale analysis, the analytical solutions are expressed as
\begin{equation}
\begin{array}{l}
\Delta_{2,xy}^{*} = - \mu\left( \frac{{\partial {u_x}}}{{\partial y}} + \frac{{\partial {u_y}}}{{\partial x}} \right) \\
+ \frac{1}{{{S_6}}}\left[ {\frac{{2\rho T}}{{{S_5}}}\frac{{D + I - 1}}{{D + I}}\frac{{\partial {u_x}}}{{\partial x}}\frac{{\partial {u_y}}}{{\partial x}} - \frac{{2\rho T}}{{{S_5}}}\frac{1}{{D + I}}\frac{{\partial {u_y}}}{{\partial x}}\frac{{\partial {u_y}}}{{\partial y}}} \right. - \frac{{2\rho T}}{{{S_6}}}\frac{1}{{D + I}}\frac{{\partial {u_x}}}{{\partial x}}\frac{{\partial {u_x}}}{{\partial y}} \\
- \frac{{2\rho T}}{{{S_6}}}\frac{1}{{D + I}}\frac{{\partial {u_x}}}{{\partial y}}\frac{{\partial {u_y}}}{{\partial y}} - \frac{{2\rho T}}{{{S_6}}}\frac{1}{{D + I}}\frac{{\partial {u_x}}}{{\partial x}}\frac{{\partial {u_y}}}{{\partial x}} - \frac{{2\rho T}}{{{S_6}}}\frac{1}{{D + I}}\frac{{\partial {u_y}}}{{\partial x}}\frac{{\partial {u_y}}}{{\partial y}} + \frac{\rho }{{{S_6}}}\frac{{2\lambda^{\prime}Q}}{{D + I}}\frac{{\partial {u_x}}}{{\partial y}} \\
+ \frac{\rho }{{{S_6}}}\frac{{2\lambda^{\prime}Q}}{{D + I}}\frac{{\partial {u_y}}}{{\partial x}} + \frac{{2{T^2}}}{{\rho {S_6}}}\frac{{\partial \rho }}{{\partial x}}\frac{{\partial \rho }}{{\partial y}} - \frac{T}{{{S_6}}}\frac{{\partial \rho }}{{\partial x}}\frac{{\partial T}}{{\partial y}} - \frac{{2{T^2}}}{{{S_6}}}\frac{{{\partial ^2}\rho }}{{\partial x\partial y}} - \frac{{2\rho T}}{{{S_6}}}\frac{{{\partial ^2}T}}{{\partial x\partial y}} \\
- \frac{T}{{{S_6}}}\frac{{\partial T}}{{\partial x}}\frac{{\partial \rho }}{{\partial y}} + \frac{{2\rho T}}{{{S_7}}}\frac{{D + I - 1}}{{D + I}}\frac{{\partial {u_x}}}{{\partial y}}\frac{{\partial {u_y}}}{{\partial y}} - \frac{{2\rho T}}{{{S_7}}}\frac{1}{{D + I}}\frac{{\partial {u_x}}}{{\partial x}}\frac{{\partial {u_x}}}{{\partial y}} + \frac{T}{{{S_{11}}}}\frac{{\partial \rho }}{{\partial x}}\frac{{\partial T}}{{\partial y}} \\
+ \frac{\rho }{{{S_{11}}}}\frac{{\partial T}}{{\partial x}}\frac{{\partial T}}{{\partial y}} + \frac{{\rho T}}{{{S_{11}}}}\frac{{{\partial ^2}T}}{{\partial x\partial y}} + \frac{T}{{{S_{12}}}}\frac{{\partial \rho }}{{\partial y}}\frac{{\partial T}}{{\partial x}} + \frac{\rho }{{{S_{12}}}}\frac{{\partial T}}{{\partial x}}\frac{{\partial T}}{{\partial y}} + \left. {\frac{{\rho T}}{{{S_{12}}}}\frac{{{\partial ^2}T}}{{\partial x\partial y}}} \right] \tt{,} \\ 
\end{array} 
\end{equation}
\begin{equation}
\begin{array}{l}
\Delta_{3,1,y}^{*} = - \kappa \frac{\partial T}{\partial y} \\
+ \frac{1}{{2{S_9}}}\left\{ { - \left( {D + I + 4} \right)\frac{{{T^2}}}{{{S_6}}}\frac{{\partial {u_x}}}{{\partial y}}\frac{{\partial \rho }}{{\partial x}} - \left( {D + I + 4} \right)\frac{{{T^2}}}{{{S_6}}}\frac{{\partial {u_y}}}{{\partial x}}\frac{{\partial \rho }}{{\partial x}}} \right. \\
- \left( {D + I + 4} \right)\frac{{\rho T}}{{{S_6}}}\frac{{\partial {u_x}}}{{\partial y}}\frac{{\partial T}}{{\partial x}} - \left( {D + I + 4} \right)\frac{{\rho T}}{{{S_6}}}\frac{{\partial {u_y}}}{{\partial x}}\frac{{\partial T}}{{\partial x}} \\
- \left( {D + I + 2} \right)\frac{{\rho {T^2}}}{{{S_6}}}\frac{{{\partial ^2}{u_x}}}{{\partial x\partial y}} - \left( {D + I + 2} \right)\frac{{\rho {T^2}}}{{{S_6}}}\frac{{{\partial ^2}{u_y}}}{{\partial x\partial x}} \\
- \left( {D + I + 4} \right)\frac{{2{T^2}}}{{{S_7}}}\frac{{D + I - 1}}{{D + I}}\frac{{\partial {u_y}}}{{\partial y}}\frac{{\partial \rho }}{{\partial y}} + \left( {D + I + 4} \right)\frac{{2{T^2}}}{{{S_7}}}\frac{1}{{D + I}}\frac{{\partial {u_x}}}{{\partial x}}\frac{{\partial \rho }}{{\partial y}} \\
- \left( {D + I + 4} \right)\frac{{2\rho T}}{{{S_7}}}\frac{{D + I - 1}}{{D + I}}\frac{{\partial {u_y}}}{{\partial y}}\frac{{\partial T}}{{\partial y}} 
+ \left( {D + I + 4} \right)\frac{{2\rho T}}{{{S_7}}}\frac{1}{{D + I}}\frac{{\partial {u_x}}}{{\partial x}}\frac{{\partial T}}{{\partial y}} \\
- \left( {D + I + 2} \right)\frac{{2\rho {T^2}}}{{{S_7}}}\frac{{D + I - 1}}{{D + I}}\frac{{{\partial ^2}{u_y}}}{{\partial y\partial y}} 
+ \left( {D + I + 2} \right)\frac{{2\rho {T^2}}}{{{S_7}}}\frac{1}{{D + I}}\frac{{{\partial ^2}{u_x}}}{{\partial x\partial y}} \\
+ \left( {D + I + 2} \right)\frac{{\rho T}}{{{S_8}}}\frac{{\partial {u_y}}}{{\partial x}}\frac{{\partial T}}{{\partial x}} + \left( {D + I + 2} \right)\frac{\rho }{{{S_9}}}\frac{{2\lambda^{\prime}Q}}{{D + I}}\frac{{\partial T}}{{\partial y}} - \frac{{4\rho T}}{{{S_9}}}\frac{{D + I + 2}}{{D + I}}\frac{{\partial {u_x}}}{{\partial x}}\frac{{\partial T}}{{\partial y}} \\
- \frac{{4\rho T}}{{{S_9}}}\frac{{D + I + 2}}{{D + I}}\frac{{\partial {u_y}}}{{\partial y}}\frac{{\partial T}}{{\partial y}} - \frac{{2\rho {T^2}}}{{{S_9}}}\frac{{D + I + 2}}{{D + I}}\frac{{{\partial ^2}{u_x}}}{{\partial x\partial y}} 
- \frac{{2\rho {T^2}}}{{{S_9}}}\frac{{D + I + 2}}{{D + I}}\frac{{{\partial ^2}{u_y}}}{{\partial y\partial y}} 
\\
- \left( {D + I + 2} \right)\frac{{\rho T}}{{{S_9}}}\frac{{\partial {u_x}}}{{\partial y}}\frac{{\partial T}}{{\partial x}} + \frac{{2\rho T}}{{{S_{11}}}}\frac{{\partial {u_x}}}{{\partial x}}\frac{{\partial T}}{{\partial y}} + \frac{{2\rho T}}{{{S_{12}}}}\frac{{\partial {u_y}}}{{\partial x}}\frac{{\partial T}}{{\partial x}} + \frac{{2\rho T}}{{{S_{12}}}}\frac{{\partial {u_x}}}{{\partial y}}\frac{{\partial T}}{{\partial x}} \\
+ \frac{{6\rho T}}{{{S_{13}}}}\frac{{\partial {u_y}}}{{\partial y}}\frac{{\partial T}}{{\partial y}} 
+ \left( {D + I + 4} \right)\frac{{{T^2}}}{{{S_{15}}}}\frac{{\partial {u_x}}}{{\partial y}}\frac{{\partial \rho }}{{\partial x}} + \left( {D + I + 4} \right)\frac{{{T^2}}}{{{S_{15}}}}\frac{{\partial {u_y}}}{{\partial x}}\frac{{\partial \rho }}{{\partial x}} \\
+ \left( {D + I + 4} \right)\frac{{2\rho T}}{{{S_{15}}}}\frac{{\partial {u_x}}}{{\partial y}}\frac{{\partial T}}{{\partial x}} + \left( {D + I + 4} \right)\frac{{2\rho T}}{{{S_{15}}}}\frac{{\partial {u_y}}}{{\partial x}}\frac{{\partial T}}{{\partial x}} 
\\
+ \left( {D + I + 4} \right)\frac{{\rho {T^2}}}{{{S_{15}}}}\frac{{{\partial ^2}{u_x}}}{{\partial x\partial y}} + \left( {D + I + 4} \right)\frac{{\rho {T^2}}}{{{S_{15}}}}\frac{{{\partial ^2}{u_y}}}{{\partial x\partial x}} \\
+ \left[ {\left( {D + I + 3} \right)\frac{{2{T^2}}}{{{S_{16}}}} - \frac{{8{T^2}}}{{{S_{16}}}}\frac{1}{{D + I}}} \right]\frac{{\partial {u_y}}}{{\partial y}}\frac{{\partial \rho }}{{\partial y}} \\
+ \left[ {\left( {D + I + 3} \right)\frac{{4\rho T}}{{{S_{16}}}} - \frac{{16\rho T}}{{{S_{16}}}}\frac{1}{{D + I}}} \right]\frac{{\partial {u_y}}}{{\partial y}}\frac{{\partial T}}{{\partial y}} \\
+ \left[ {\left( {D + I + 3} \right)\frac{{2\rho {T^2}}}{{{S_{16}}}} - \frac{{8\rho {T^2}}}{{{S_{16}}}}\frac{1}{{D + I}}} \right]\frac{{{\partial ^2}{u_y}}}{{\partial y\partial y}} \\
- \frac{{2{T^2}}}{{{S_{16}}}}\frac{{D + I + 4}}{{D + I}}\frac{{\partial {u_x}}}{{\partial x}}\frac{{\partial \rho }}{{\partial y}} - \frac{{4\rho T}}{{{S_{16}}}}\frac{{D + I + 4}}{{D + I}}\frac{{\partial {u_x}}}{{\partial x}}\frac{{\partial T}}{{\partial y}} \left. { - \frac{{2\rho {T^2}}}{{{S_{16}}}}\frac{{D + I + 4}}{{D + I}}\frac{{{\partial ^2}{u_x}}}{{\partial x\partial y}}} \right\} \tt{.}\\ 
\end{array} 
\end{equation}
The CDBM results are in excellent agreement with the analytic solutions. This demonstrates that the CDBM can accurately describe nonequilibrium behavior.

\subsection{Homogeneous chemical reaction system}

Now, to verify that the present model can naturally couple the chemical reactions with physical fields, we consider a homogeneous chemical reaction system, where the physical field is uniformly distributed in a closed space. The chemical reactants are uniformly mixed and distributed statically within this space. When the reactants reach the reaction conditions, the chemical reaction begins, generating products and releasing heat, i.e., converting chemical energy into the system's internal energy. Upon completion of the reaction, all chemical energy is released into the system's internal energy, resulting in a rise in temperature. 
Initially, the system is at rest with the temperature $T_0=2.0$. 
Upon completion of the chemical reaction, the system's temperature is related to the chemical heat release per unit mass of fuel $Q$, and the additional degree of freedom $I$, as given by the following equation:
\begin{equation}
T = {T_0} + \frac{{2Q}}{{D + I}}
\tt{.}
\label{T}
\end{equation}
In this simulation, the time step is $\Delta t = 2.0 \times 10^{-6}$, the spatial step $\Delta x = \Delta y = 1.0 \times 10^{-5}$, the relaxation time $\tau = 1.0 \times 10^{-5}$, with discrete parameters $v_a = 1.3$, $v_b = 2.7$, $v_c = 3.3$, $\eta_a = 1.2$, $\eta_b = 0.5$, and $\eta_c = 5.0$. Since the system is homogeneous and the physical quantities are uniformly distributed, only one grid point is used for computational efficiency, i.e., $N_x \times N_y = 1 \times 1$. Additionally, periodic boundary conditions are applied to all sides of the computational domain.

\begin{figure}
	\begin{center}
		\includegraphics[width=0.45\textwidth]{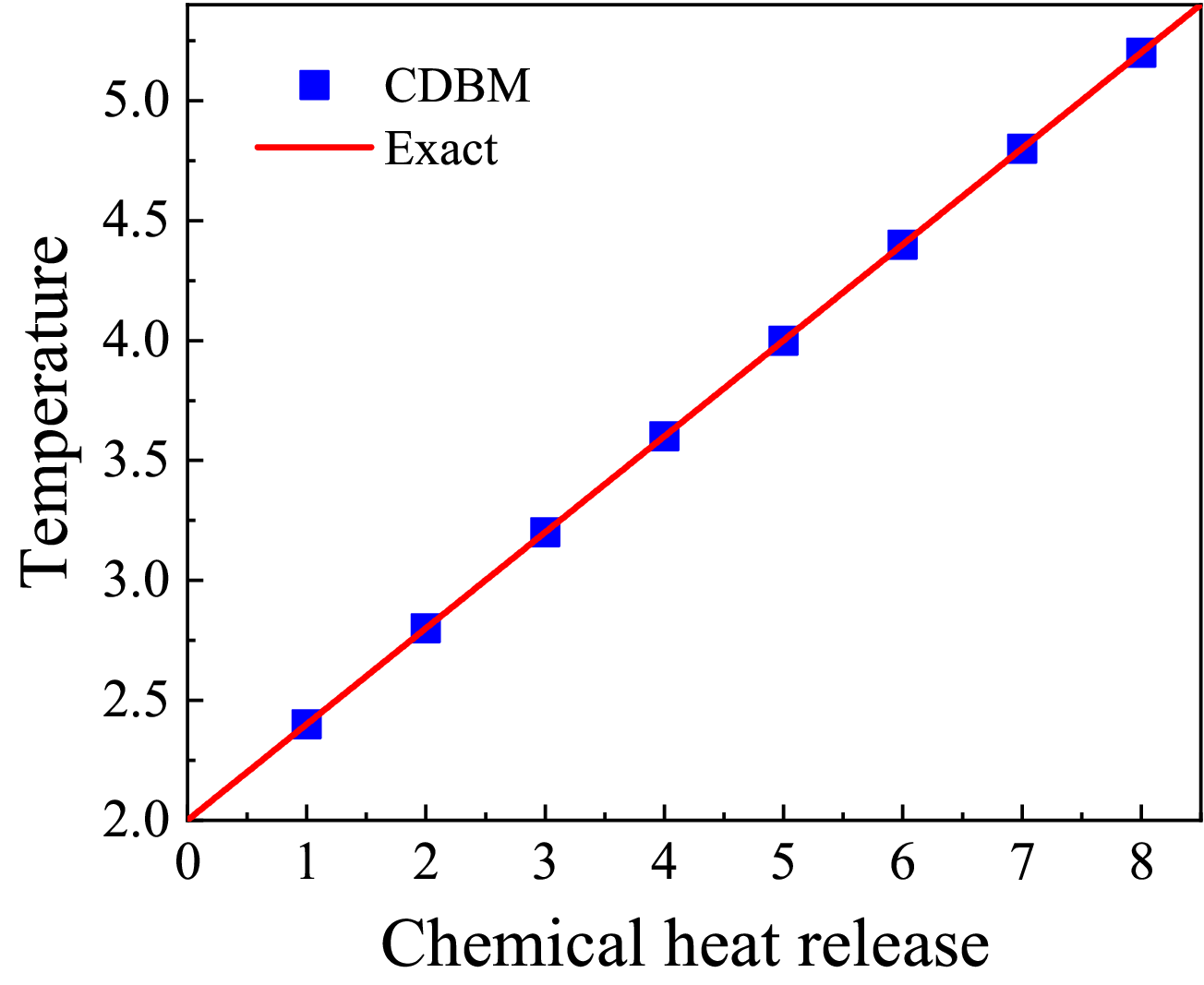}
	\end{center}
	\setlength{\abovecaptionskip}{-0.4cm}
	\caption{Temperature after chemical reaction under various
		chemical heat release with $\gamma=1.4$.}
	\label{Fig08}
\end{figure}

Figure \ref{Fig08} illustrates the temperature resulting from various chemical heat release per unit mass of fuel that undergoes complete reaction. The corresponding relationship is given by Eq. (\ref{T}), where the specific heat ratio $\gamma=1.4$ is set. The symbols indicate the CDBM simulation results while the solid line represents the analytical solutions. It can be seen that the CDBM simulation results are in perfect agreement with the analytical solutions. This demonstrates that the model is suitable for chemical reacting flow systems with a wide range of reaction heats.

\begin{figure}
	\begin{center}
		\includegraphics[width=0.45\textwidth]{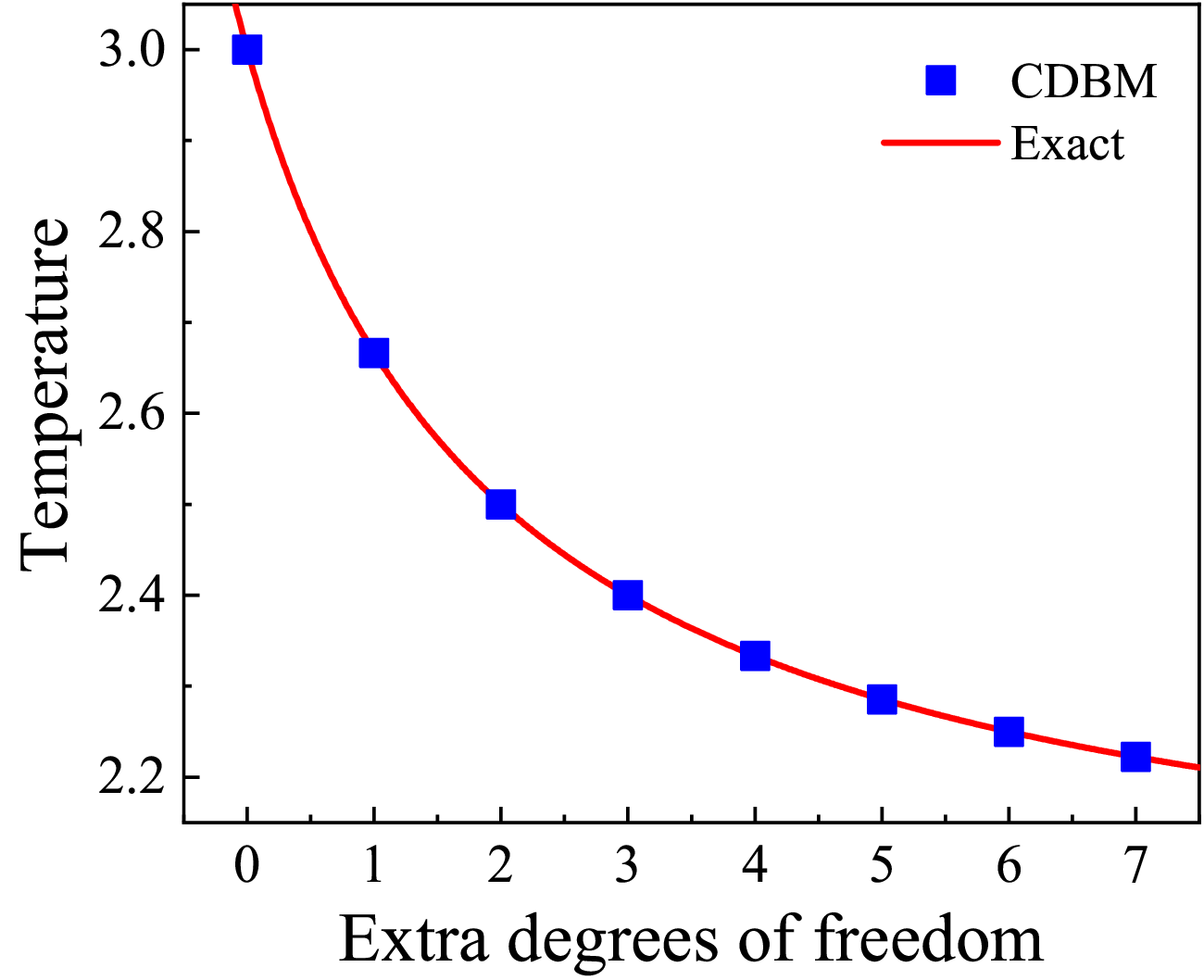}
	\end{center}
	\setlength{\abovecaptionskip}{-0.4cm}
	\caption{Temperature after chemical reaction under various
		extra degrees of freedom with $Q=1.0$.}
	\label{Fig09}
\end{figure}

Next, we consider the case where the chemical energy constant is fixed at $Q = 1.0$, and the extra degrees of freedom are varied. Figure \ref{Fig09} illustrates the system's temperature corresponding to different extra degrees of freedom after the chemical reaction. Symbols represent the CDBM simulation results, and the solid line shows the theoretical solutions in Eq. (\ref{T}). Similarly, the CDBM results closely match the analytical solutions. These results further demonstrate the model's applicability to reacting flow systems with varying extra degrees of freedom.

\subsection{Detonation wave}

A detonation wave is a distinct reactive flow phenomenon characterized by rapid increases in density, temperature, and pressure, accompanied by drastic changes in the physical field during the evolution of the wave. This section verifies that the CDBM can effectively simulate the propagation of a detonation wave, using a two-step reaction scheme to model the chemical kinetics \cite{Ng2005numerical}. The detonation wave propagates in the horizontal direction, with the initial physical conditions defined as follows:
\begin{equation}
\left\{ \begin{array}{l}
{\left( {\rho ,{u_x},{u_y},T,\lambda } \right)_L} = \left( {1.3884,0.5774,0.0,1.5786,1.0} \right) \tt{,} \hfill \\
{\left( {\rho ,{u_x},{u_y},T,\lambda } \right)_R} = \left( {1.0,0.0,0.0,1.0,0.0} \right) \tt{.} \hfill \\
\end{array}  \right.
\end{equation}
The subscripts $L$ indicates $0 \leq x \leq 0.05$ and $R$ indicates $0.05 \leq x \leq 0.5$. The physical states satisfy the Rankine--Hugoniot condition across the interface. The reaction heat is set to $Q = 1.0$, with the Mach number Ma $= 1.74$, specific heat ratio $\gamma = 1.4$, and relaxation parameter $S_i = 10^{5}$. The discrete velocity and internal energy parameters are ${v_a} = 1.3$, ${v_b} = 2.7$, ${v_c} = 3.3$, ${\eta_a} = 1.2$, ${\eta_b} = 0.5$, and ${\eta_c} = 5.0$. Additional numerical parameters include the time step $\Delta t = 2.0 \times 10^{-6}$, spatial step $\Delta x = \Delta y = 1.0 \times 10^{-5}$, and a grid resolution of $N_x \times N_y = 50000 \times 1$. Inflow/outflow boundary conditions are applied in the $x$-direction, and periodic conditions are imposed in the $y$-direction.

\begin{figure}
	\begin{center}
		\includegraphics[width=1.0\textwidth]{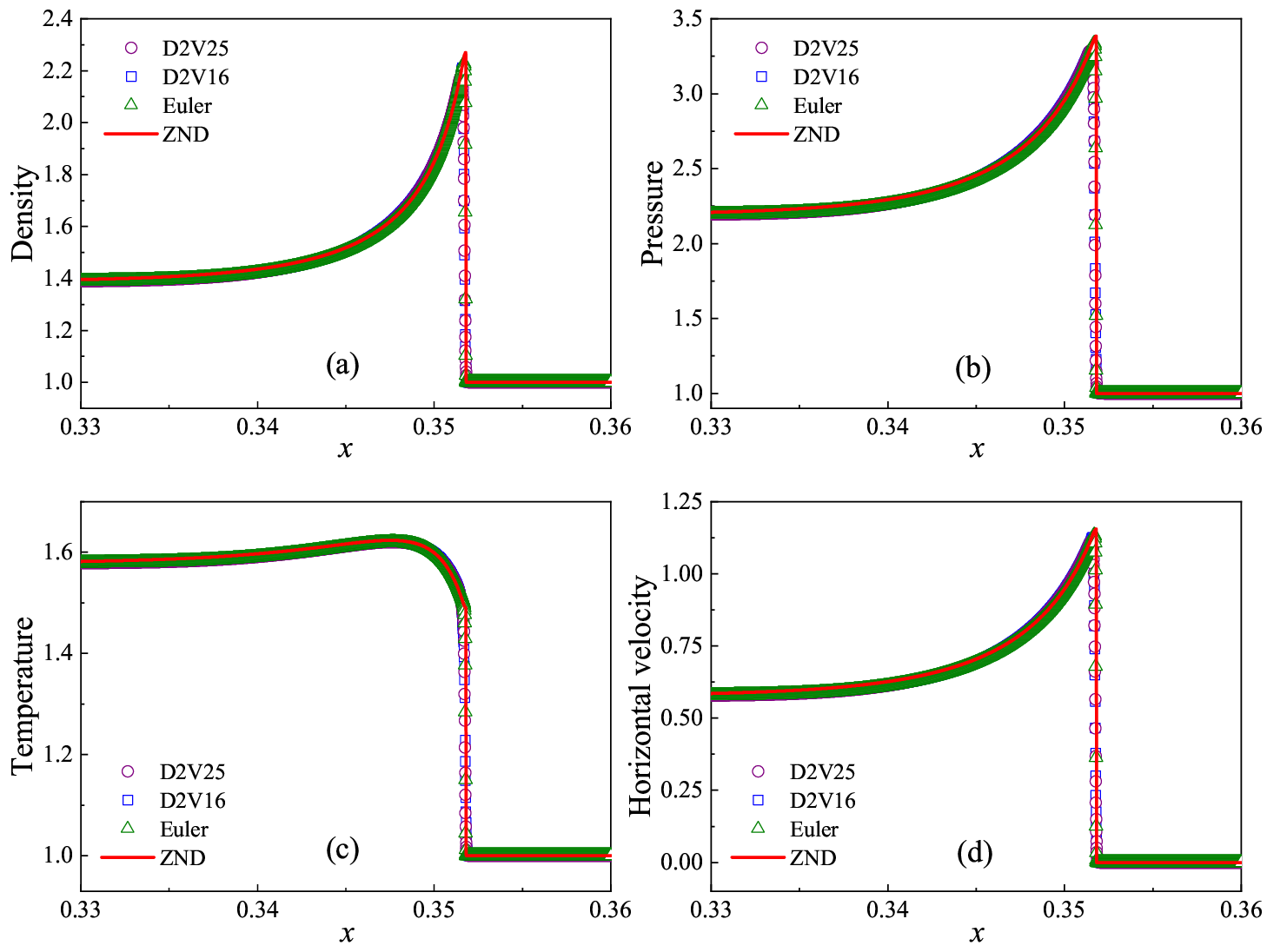}
	\end{center}
	\setlength{\abovecaptionskip}{-0.4cm}
	\caption{Distribution of physical quantities in the vicinity of the detonation wave: (a) density, (b) pressure, (c) temperature, and (d) horizontal velocity. Circles, squares, triangles, and solid lines represent the results of D2V25, D2V16, Euler solver, and ZND theory, respectively.}
	\label{Fig10}
\end{figure}

Figure \ref{Fig10} depicts the profiles of density (a), pressure (b), temperature (c), and horizontal velocity (d) at $t = 0.15$ during the detonation. Circles, squares, and triangles represent results from the D2V25, D2V16, and Euler solver, respectively. Solid lines indicate the solutions of Zel'dovich--Neumann--D$\ddot{\rm{o}}$ring (ZND) theory \cite{Law2006}. It can be found that the results of the Euler solver are in perfect agreement with the ZND solutions in the whole region. This consistency arises from the fact that both methods neglect viscosity, heat conduction, and other nonequilibrium effects, and idealize the detonation wave as a strong discontinuity. In contrast, the CDBM based on the D2V16 and D2V25 models incorporates viscosity, heat conduction, and essential nonequilibrium effects, resulting in noticeable deviations near the von Neumann peak. Notably, the D2V25 model more accurately capture the underlying physical processes compared to the D2V16 model, thus its predictions exhibit slightly larger deviations from the ZND solutions. Physically, the D2V25 model takes more nonequilibrium effects into account and offers higher fidelity, which naturally leads to a departure from the idealized assumptions underlying the ZND theory.
To be specific, the post-shock state predicted by the CDBM with D2V25 is $\left(\rho,u_x,u_y,T\right) = \left(1.3894,0.5777,0,1.5770\right)$, while the ZND yields $\left(1.3884,0.5774,0,2.1916,1.5786\right)$. The corresponding relative errors are $0.07\%$, $0.05\%$, $0\%$, $0.02\%$, and $0.1\%$, respectively. These results indicate that the CDBM is capable of accurately capturing the physical mechanisms overlooked by idealized models, while maintaining a high level of numerical accuracy.

\begin{figure}
	\begin{center}
		\includegraphics[width=0.5\textwidth]{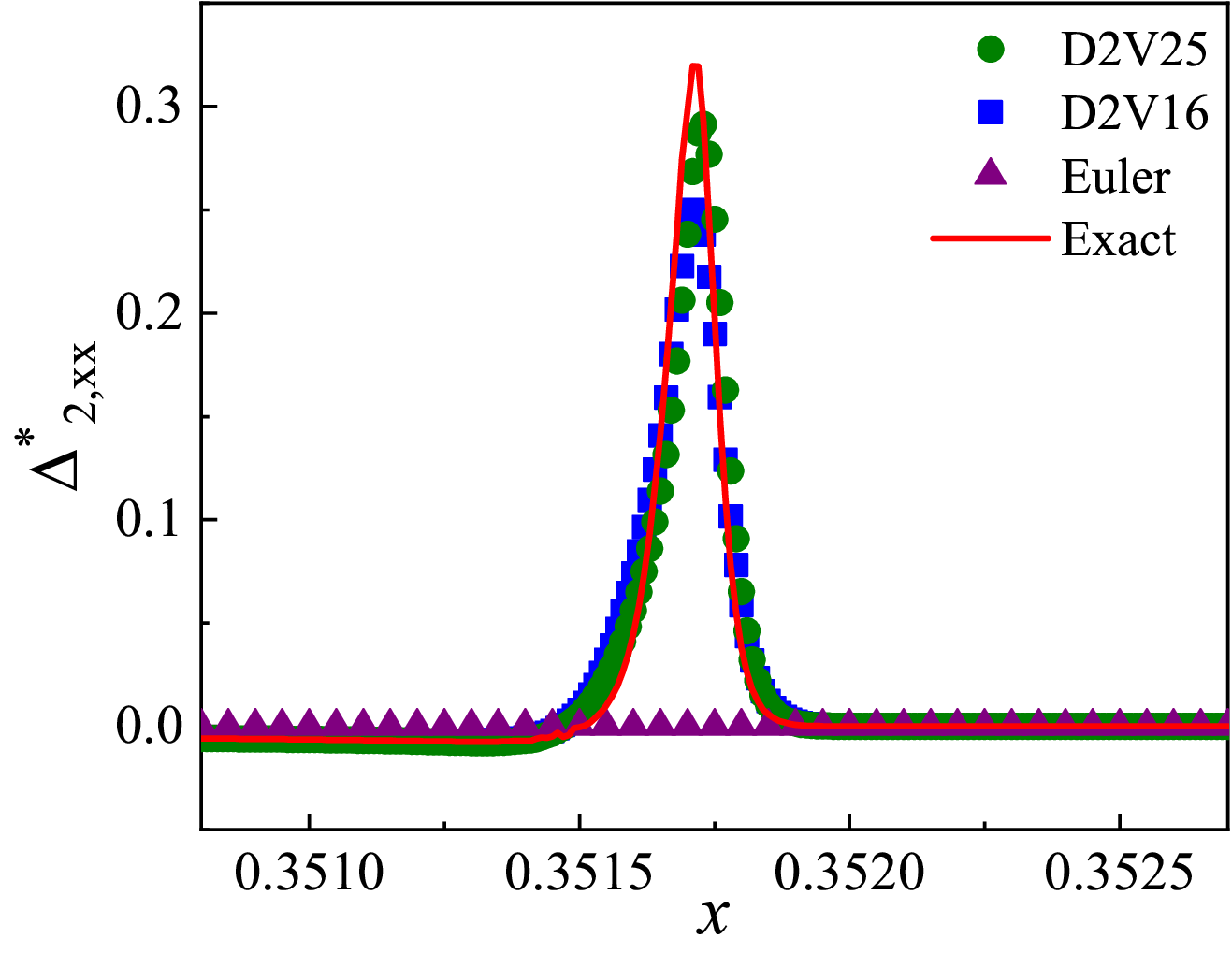}
	\end{center}
	\setlength{\abovecaptionskip}{-0.4cm}
	\caption{Profiles of the nonequilibrium manifestation $\Delta_{2,xx}^{*}$ around the detonation wave. The circle, square, triangle, and solid line correspond to the D2V25, D2V16, Euler solver, and theory, respectively.}
	\label{Fig11}
\end{figure}

Figure \ref{Fig11} illustrates the nonequilibrium quantity $\Delta_{2,xx}^{*}$ versus $x$ at the time instant $t=0.15$ in the evolution of the detonation wave. Symbols denote the numerical results of various models, and the solid line represent the following exact solutions:
\begin{equation}
\begin{array}{l}
\Delta_{2,xx}^{*} = - \frac{{2\rho T}}{{S_5}}\left( \frac{D + I - 1}{D + I}\frac{\partial {u_x}}{\partial x} - \frac{1}{D + I}\frac{\partial {u_y}}{\partial y} \right)\\
+ \frac{1}{{{S_5}}}\left[ {\left( {\frac{{2\rho T}}{{{S_5}}}\frac{{D + I - 1}}{{D + I}} - \frac{{8\rho T}}{{{S_5}}}\frac{{D + I - 1}}{{{{\left( {D + I} \right)}^2}}}} \right)\frac{{\partial {u_x}}}{{\partial x}}\frac{{\partial {u_x}}}{{\partial x}}} \right. \\
+ \left( {\frac{{2\rho T}}{{{S_5}}}\frac{1}{{D + I}} + \frac{{4\rho T}}{{{S_5}}}\frac{1}{{{{\left( {D + I} \right)}^2}}}} \right)\frac{{\partial {u_y}}}{{\partial y}}\frac{{\partial {u_y}}}{{\partial y}} \\
+ \left( { - \frac{{8\rho T}}{{{S_5}}}\frac{1}{{D + I}} + \frac{{12\rho T}}{{{S_5}}}\frac{1}{{{{\left( {D + I} \right)}^2}}}} \right)\frac{{\partial {u_x}}}{{\partial x}}\frac{{\partial {u_y}}}{{\partial y}} + \frac{{2\rho }}{{{S_5}}}\frac{{D + I - 1}}{{D + I}}\frac{{2\lambda ^{\prime}Q}}{{D + I}}\frac{{\partial {u_x}}}{{\partial x}} \\
- \frac{{2\rho }}{{{S_5}}}\frac{1}{{D + I}}\frac{{2\lambda^{\prime}Q}}{{D + I}}\frac{{\partial {u_y}}}{{\partial y}} + \frac{{2{T^2}}}{{{S_5}}}\frac{{D + I - 1}}{{\rho \left( {D + I} \right)}}\frac{{\partial \rho }}{{\partial x}}\frac{{\partial \rho }}{{\partial x}} - \frac{{2T}}{{{S_5}}}\frac{{D + I - 1}}{{D + I}}\frac{{\partial T}}{{\partial x}}\frac{{\partial \rho }}{{\partial x}} \\
- \frac{{2{T^2}}}{{{S_5}}}\frac{{D + I - 1}}{{D + I}}\frac{{{\partial ^2}\rho }}{{\partial x\partial x}} - \frac{{2\rho T}}{{{S_5}}}\frac{{D + I - 1}}{{D + I}}\frac{{{\partial ^2}T}}{{\partial x\partial x}} + \left( { - \frac{{2\rho T}}{{{S_5}}} + \frac{{4\rho T}}{{{S_5}}}\frac{1}{{D + I}}} \right)\frac{{\partial {u_x}}}{{\partial y}}\frac{{\partial {u_y}}}{{\partial x}} \\
- \frac{{2{T^2}}}{{{S_5}}}\frac{1}{{\rho \left( {D + I} \right)}}\frac{{\partial \rho }}{{\partial y}}\frac{{\partial \rho }}{{\partial y}} + \frac{{2T}}{{{S_5}}}\frac{1}{{D + I}}\frac{{\partial T}}{{\partial y}}\frac{{\partial \rho }}{{\partial y}} 
+ \frac{{2{T^2}}}{{{S_5}}}\frac{1}{{D + I}}\frac{{{\partial ^2}\rho }}{{\partial y\partial y}} 
+ \frac{{2\rho T}}{{{S_5}}}\frac{1}{{D + I}}\frac{{{\partial ^2}T}}{{\partial y\partial y}} \\
+ \frac{{2\rho T}}{{{S_6}}}\frac{{D + I - 1}}{{D + I}}\frac{{\partial {u_x}}}{{\partial y}}\frac{{\partial {u_x}}}{{\partial y}} + \frac{{2\rho T}}{{{S_6}}}\frac{{D + I - 2}}{{D + I}}\frac{{\partial {u_x}}}{{\partial y}}\frac{{\partial {u_y}}}{{\partial x}} - \frac{{2\rho T}}{{{S_6}}}\frac{1}{{D + I}}\frac{{\partial {u_y}}}{{\partial x}}\frac{{\partial {u_y}}}{{\partial x}} \\
- \frac{{4\rho T}}{{{S_7}}}\frac{{D + I - 1}}{{{{\left( {D + I} \right)}^2}}}\frac{{\partial {u_y}}}{{\partial y}}\frac{{\partial {u_y}}}{{\partial y}} + \frac{{4\rho T}}{{{S_7}}}\frac{1}{{{{\left( {D + I} \right)}^2}}}\frac{{\partial {u_x}}}{{\partial x}}\frac{{\partial {u_y}}}{{\partial y}} - \frac{T}{{{S_8}}}\frac{{D + I + 2}}{{D + I}}\frac{{\partial T}}{{\partial x}}\frac{{\partial \rho }}{{\partial x}} \\
- \frac{\rho }{{{S_8}}}\frac{{D + I + 2}}{{D + I}}\frac{{\partial T}}{{\partial x}}\frac{{\partial T}}{{\partial x}} - \frac{{\rho T}}{{{S_8}}}\frac{{D + I + 2}}{{D + I}}\frac{{{\partial ^2}T}}{{\partial x\partial x}} - \frac{T}{{{S_9}}}\frac{{D + I + 2}}{{D + I}}\frac{{\partial T}}{{\partial y}}\frac{{\partial \rho }}{{\partial y}} - \frac{\rho }{{{S_9}}}\frac{{D + I + 2}}{{D + I}}\frac{{\partial T}}{{\partial y}}\frac{{\partial T}}{{\partial y}} \\
- \frac{{\rho T}}{{{S_9}}}\frac{{D + I + 2}}{{D + I}}\frac{{{\partial ^2}T}}{{\partial y\partial y}} + \frac{{3T}}{{{S_{10}}}}\frac{{\partial T}}{{\partial x}}\frac{{\partial \rho }}{{\partial x}} + \frac{{3\rho }}{{{S_{10}}}}\frac{{\partial T}}{{\partial x}}\frac{{\partial T}}{{\partial x}} + \frac{{3\rho T}}{{{S_{10}}}}\frac{{{\partial ^2}T}}{{\partial x\partial x}} + \frac{T}{{{S_{11}}}}\frac{{\partial T}}{{\partial y}}\frac{{\partial \rho }}{{\partial y}} \\
+ \frac{\rho }{{{S_{11}}}}\frac{{\partial T}}{{\partial y}}\frac{{\partial T}}{{\partial y}} + \left. {\frac{{\rho T}}{{{S_{11}}}}\frac{{{\partial ^2}T}}{{\partial y\partial y}}} \right] \tt{.} \\ 
\end{array} 
\label{delta_2xx^*}
\end{equation}
It can be found that D2V25 is in better agreement with the analytic solution than D2V16. Since the Euler solver ignores viscosity and heat conduction, the corresponding nonequilibrium effects remain zero. It is demonstrated that the CDBM with D2V25 can extract and characterize the nonequilibrium effects accurately.

\subsection{Collision of two detonation waves}

\begin{figure}
	\begin{center}
		\includegraphics[width=0.5\textwidth]{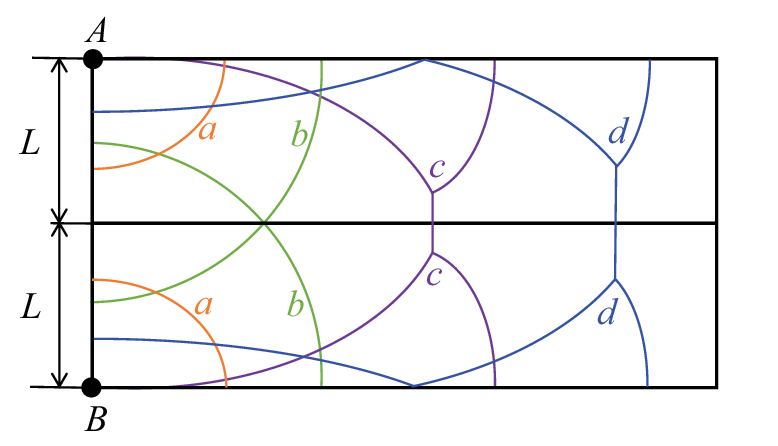}
	\end{center}
	\setlength{\abovecaptionskip}{-0.4cm}
	\caption{Schematic diagram of the collision of two detonation waves.}
	\label{Fig12}
\end{figure}

Finally, let us validate that the proposed model can accurately capture the collision of two detonation waves in the two-dimensional system, which is a challenging test. Figure \ref{Fig12} illustrates a rectangular computational domain, where the combustible mixture is uniformly distributed, with points $A$ and $B$ separated by $2L$. Ignition occurs at the symmetric points $A$ and $B$. In the early stage, two symmetric detonation waves, labeled by ``$a$", are subsequently generated. Over time, the two waves collide, and then form the waves labeled by ``$b$". After a period of interaction, the Mach reflection occurs, which can be found in curves ``$c$" and ``$d$". To simulate the above process, the initial configuration is set as follows:
\begin{equation}
\left( {\rho ,\mathbf{u},T} \right) = \left\{ \begin{array}{l}
\left( {1.0,0.0,2.0} \right),{\text{ at points }}A{\text{ and }}B \tt{,} \hfill \\
\left( {1.0,0.0,1.0} \right),{\text{ else. }} \hfill \\
\end{array}  \right.
\end{equation}
The heat of reaction is $Q=2.0$, the relaxation parameter $S_i=10^{5}$, the grid size ${N_x} \times {N_y} = 600 \times 400$, the time step $\Delta t = 1.0 \times {10^{-5}}$, the spatial step $\Delta x = \Delta y = 1.0 \times {10^{-3}}$. In addition, an extrapolated boundary condition is used for the right boundary, and specular reflection boundary conditions are for other boundaries. 

\begin{figure}
	\begin{center}
		\includegraphics[width=0.93\textwidth]{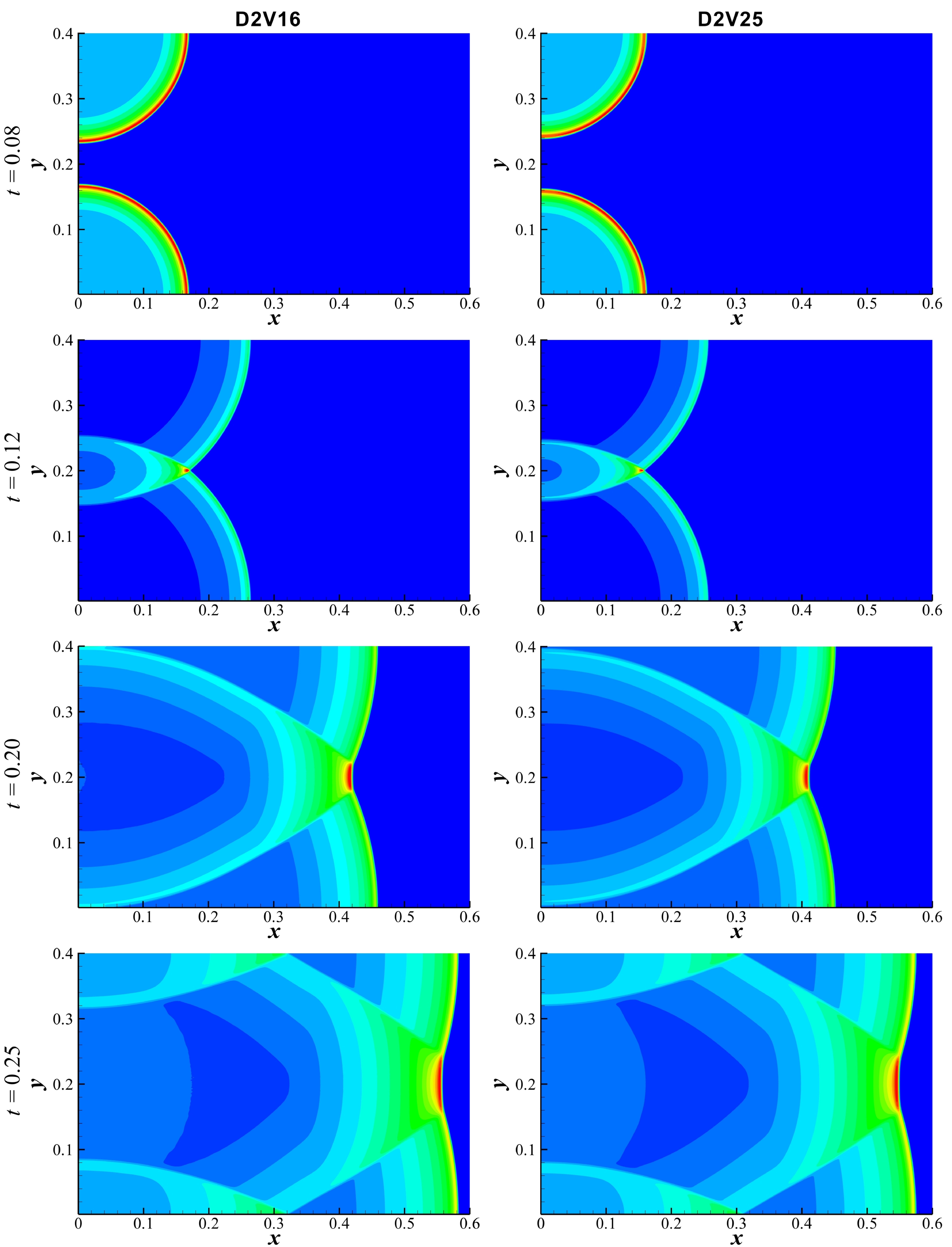}
	\end{center}
	\setlength{\abovecaptionskip}{-0.4cm}
	\caption{The process of the collision of two detonation waves simulated by the D2V16 (left) and the D2V25 (right). The color gradient from blue to red indicates an increase in pressure.}
	\label{Fig13}
\end{figure}

Figure \ref{Fig13} illustrates the pressure contours during the collision of two detonation waves. The left and right columns in Fig. \ref{Fig13} show the results of the D2V16 and D2V25 models, respectively. The colors from blue to red indicate an increase in pressure. It can be observed that a set of symmetric detonation waves are generated after ignition, then propagate and collide at the symmetry plane. After a period of time, Mach stem is generated where the pressure is extremely high. As time evolves, the transverse wave behind the wave collides with and reflects off the upper and lower walls. Three focal points appear near the interaction point of the detonation waves. 

It can be found in Fig. \ref{Fig13} that both D2V16 and D2V25 can accurately describe the collision of two detonation waves, including the whole process of ignition, propagation, collision, and reflection. Particularly, the generation of Mach stems and triple points are captured at the time instants $t=0.20$ and $t=0.25$. Moreover, the similarity of their results proves the accuracy in modeling two-dimensional detonation waves. Subtle variations in the pressure profiles reveal small differences in their treatment of nonequilibrium effects. These deviations stem from the fact that, in the continuum limit, the CDBM with D2V16 recovers the NS equations, while the CDBM with D2V25 recovers the Burnett equations. Overall, these findings confirm the reliability and robustness of the CDBM in modeling systems containing chemically reactive flows.

\section{Conclusion}\label{SecIV}

In this work, a CDBM is tailored for subsonic and supersonic compressible reactive flows exhibiting both HNE and TNE effects. This model incorporates chemical reactions and allows for adjustable specific heat ratios and Prandtl numbers. A set of two-dimensional twenty-five velocities with spatial symmetry is designed for the CDBM. The matrix reversion method is employed to calculate the discrete equilibrium distribution function, as well as the collision and reaction terms. Unlike previous MRT-DBMs, the CDBM utilizes the central-moment MRT collision operator \cite{Su2025central} and requires no correction terms. 
Through CE multiscale analysis, it is demonstrated that the CDBM can recover the Burnett equations in the hydrodynamic limit. Moreover, the CDBM is capable of directly quantifying nonequilibrium effects related to thermal fluctuations, which is an improvement over the traditional DBMs based on raw moment space formulations. 

For the sake of verification and validation, several benchmarks are adopted as simulation tests. First, the thermal Couette flow is used to verify the model's capability to adjust both specific heat ratios and Prandtl numbers. Next, a homogeneous chemical reaction system demonstrates the model's ability to naturally couple chemical reactions with the physical field and its applicability to reactive flow systems with varying chemical heat releases and extra degrees of freedom. Additionally, the simulation of a steady detonation wave confirms the model's suitability for supersonic compressible reactive flows. Finally, the model successfully simulates a classical two-dimensional phenomenon involving the collision of two detonation waves.

\section*{Novelty and significance statement}

This work proposes a Burnett-level CDBM for compressible reactive flows, enabling accurate modeling of high-order hydrodynamic and thermodynamic nonequilibrium effects. The development of a discrete velocity model with enhanced spatial symmetry significantly improves numerical stability, efficiency, and isotropy. The approach provides a powerful and flexible tool for exploring complex reactive flow physics beyond the NS level, which is critical for advancing the predictive capability of combustion modeling in high-speed and high-gradient regimes. The method's theoretical depth, combined with its computational advantages, marks a notable advancement in the mesoscopic kinetic modeling of reactive flows.

\section*{CRediT authorship contribution statement}
\textbf{Qingbin Wu:} Writing -- original draft, Visualization, Validation, Methodology, Investigation, Data curation, Formal analysis. 
\textbf{Chuandong Lin:} Writing -- review \& editing, Methodology, Software, Conceptualization, Funding acquisition. 
\textbf{Huilin Lai:} Writing -- review \& editing, Supervision, Resources, Project administration, Funding acquisition.

\section*{Acknowledgment}
This work is supported by National Natural Science Foundation of China (under Grant No. U2242214), Guangdong Basic and Applied Basic Research Foundation (under Grant No. 2024A1515010927), Humanities and Social Science Foundation of the Ministry of Education in China (under Grant No. 24YJCZH163), Fujian Provincial Units Special Funds for Education and Research (K3-949), and Fundamental Research Funds for the Central Universities, Sun Yat-sen University (under Grant No. 24qnpy044). This work is partly supported by the Open Research Fund of Key Laboratory of Analytical Mathematics and Applications (Fujian Normal University), Ministry of Education, P. R. China (under Grant No. JAM2405).

\appendix

\bibliography{References}

\end{document}